\documentclass[10pt,a4paper]{article}
\usepackage{amsmath}
\usepackage{amssymb}
\usepackage{array}
\usepackage{calc}
\usepackage{longtable}
\usepackage{multirow,booktabs}
\usepackage{cite,mcite}
\usepackage{relsize}
\usepackage{graphicx}
\usepackage{xspace}
\usepackage{units}
\usepackage{afterpage}
\usepackage{placeins}

\numberwithin{equation}{section}
\usepackage{mciteplus}
\usepackage[pdfborder={0 0 0}]{hyperref}
\usepackage[format=hang,labelfont=bf,hypcap=true]{caption}
\usepackage{subcaption}
\usepackage{sectsty}
\allsectionsfont{\sffamily}
\subsubsectionfont{\mdseries\itshape\large}
\makeatletter
\DeclareRobustCommand*{\bfseries}{%
  \not@math@alphabet\bfseries\mathbf
  \fontseries\bfdefault\selectfont
  \boldmath
}
\makeatother
\let\spreprint\empty
\newcommand{\preprint}[1]{\def\spreprint{\protect#1}}
\let\sinstitute\empty

\makeatletter
\renewcommand{\maketitle}{\begingroup
  \null\thispagestyle{empty}%
    \ifx\spreprint\empty
      \vskip 5ex
    \else
      \flushright\large\spreprint\vskip 2ex
    \fi
    \vskip 5ex
    \flushleft
      {\sffamily\bfseries\huge\@title}\vskip 6ex
      \@author\vskip 2ex
      \ifx\sinstitute\empty
      \else
        {\small\sinstitute}
      \fi
    \vskip 5ex
  \endgroup
}
\makeatother
\renewenvironment{abstract}{\begin{center}
  {\large\sffamily\bfseries Abstract: }
  \begin{minipage}[t]{0.75\textwidth}
}{\end{minipage}\end{center}\vskip 10ex}

\numberwithin{equation}{section}

\newcommand{\bea}{\begin{eqnarray}}
\newcommand{\eea}{\end{eqnarray}}
\newcommand{\beq}{\begin{equation}}
\newcommand{\eeq}{\end{equation}}

\newcommand{\bs}{\begin{split}}
\newcommand{\es}{\end{split}}

\newcommand{\bi}{\begin{itemize}}
\newcommand{\ei}{\end{itemize}}
\newcommand{\bc}{\begin{center}}
\newcommand{\ec}{\end{center}}
\newcommand{\bac}{\begin{array}{c}}
\newcommand{\bacc}{\begin{array}{cc}}
\newcommand{\ea}{\end{array}}

\def\spa#1.#2{\langle#1\,#2\rangle}
\def\spb#1.#2{[#1\,#2]}

\def\d{{\rm d}}

\newcommand{\UGeV}{\ensuremath{\,\mathrm{GeV}}\xspace}

\newcommand{\sla}[1]{\ensuremath{{#1\kern-0.45em/}}}

\newcommand\HERA{H\scalebox{0.8}{ERA}\xspace}
\newcommand\EIC{E\scalebox{0.8}{IC}\xspace}

\newcommand\lhc{L\protect\scalebox{0.8}{HC}\xspace}
\newcommand\LHC{L\protect\scalebox{0.8}{HC}\xspace}

\newcommand\hera{H\scalebox{0.8}{ERA}\xspace}
\newcommand\hone{H1\xspace}
\newcommand\zeus{Z\scalebox{0.8}{EUS}\xspace}
\newcommand\ZEUS{\zeus}

\newcommand{\MCatNLO}{M\protect\scalebox{0.8}{C}@N\protect\scalebox{0.8}{LO}\xspace}

\newcommand{\OpenLoops}{O\protect\scalebox{0.8}{PEN}L\protect\scalebox{0.8}{OOPS}\xspace}

\newcommand{\Sherpa}{S\protect\scalebox{0.8}{HERPA}\xspace}
\newcommand{\Comix}{C\protect\scalebox{0.8}{OMIX}\xspace}

\newcommand{\Amegic}{A\protect\scalebox{0.8}{MEGIC}\xspace}

\newcommand{\Ahadic}{A\protect\scalebox{0.8}{HADIC}\xspace}

\newcommand{\Rivet}{R\protect\scalebox{0.8}{IVET}\xspace}

\newcommand{\LHAPDF}{L\protect\scalebox{0.8}{HAPDF}\xspace}

\usepackage{authblk}
\usepackage[symbol]{footmisc}
\usepackage{todonotes}

\newcommand{\Pom}{{I\!\!P}}
\newcommand{\Reg}{{I\!\!R}}

\hypersetup{
  pdfauthor={Frank Krauss, Peter Meinzinger}
  pdftitle={Hard Diffraction in Sherpa},
  pdfkeywords={QCD, Event Generators, Diffraction, HERA}
}
\begin{document}
\preprint{IPPP/24/39, MCNET-24-12}
\title{Hard Diffraction in \protect\Sherpa}
\author[1]{Frank~Krauss\footnote[1]{frank.krauss@durham.ac.uk}}
\author[1]{Peter~Meinzinger\footnote[2]{peter.meinzinger@durham.ac.uk}}
\affil[1]{Institute for Particle Physics Phenomenology, Durham University, Durham DH1 3LE, UK}

\maketitle
\begin{abstract}
We present the first complete simulation framework for the simulation of jet production in diffractive events at next-to leading order in QCD, matched to the parton shower.
We validate the implementation in the \Sherpa event generator with data from the \hone and \zeus experiments for diffractive DIS and diffractive photoproduction.
For the latter, we review different models aiming to explain the observed factorisation breaking and we argue that at NLO the direct component must also be suppressed.
We provide predictions for diffractive jet production both in DIS and in photoproduction events for the upcoming \EIC.
\end{abstract}

\section{Introduction}\label{Sec:Intro}
Diffractive events constitute a significant part of the cross section at electron-hadron colliders such as \HERA in the past and the upcoming \EIC and at hadron-hadron colliders such as the currently operating \LHC.
Especially the production of jets in diffractive events also offers potential additional insights into the dynamics of the strong interaction in its interplay between the hard perturbative scales of jet production and the soft non-perturbative scales of hadronic structures~\cite{Newman:2013ada}.
In this publication we introduce the first hadron-level simulation of diffractive jet-production at next-to leading order accuracy in the perturbative expansion of QCD at electron-hadron colliders.
At such experiments, the interaction of incident electrons and protons is mediated by a virtual photon; in dependence of its virtual mass $q^2 = -Q^2$ the interaction will either be classified as electro-production (at large $Q^2$, in the following somewhat misnamed as deep-inelastic scattering, DIS), or as photo-production (at small $Q^2$).
In the latter the photon may fluctuate into a system with hadronic structure, like, for example, a vector meson, and, correspondingly it must be described by a photon parton distribution function (PDF).
We will therefore discuss the simulation of diffractive events in both event classes and we will show results for both.

In general, the description of hard processes involving incident hadrons relies on the factorisation of the cross section calculation into PDFs which encode the transition of the hadrons into the partons, their constituents which are usually identified with the quanta of the strong interaction, quarks and gluons, and into the parton-level cross sections.
This factorization is necessary to allow systematic calculations of cross sections from first principles in quantum field theory, and their systematic improvement through perturbative expansions in the limit of small couplings, which act as expansion parameter.
This factorization picture has been proven to hold true for DIS~\cite{Collins:1989gx}, and has been a cornerstone of QCD theory ever since.
The proof was extended for diffractive DIS~\cite{Collins:1997sr}, underpinning calculations for di-jet production cross sections at next-to (NLO)~\cite{H1:2007jtx} and next-to--\-next-to leading order (NNLO) in QCD~\cite{Britzger:2018zvv}.
Diffractive PDFs were extracted at NLO accuracy mainly from \HERA data in~\cite{H1:2006zyl}; it is customary to factorise them into the flux of a pomeron (or reggeon) and the PDF of this intermediate "particle"~\cite{Ingelman:1984ns}.

However successful for the description of diffractive DIS, though, it appears as if the factorization assumption does not extrapolate to the case of diffractive photo-production events.
There are a number of underlying physics effects which may render factorization invalid, including, for example, (i) additional scattering between the hadronic structure of the photon and the proton rump, broadly speaking part of the unitarisation of the hard diffractive cross section~\cite{Kaidalov:2003gy,Kaidalov:2003xf}; (ii) the impact of hadronization effects which may impact on the emergence and survival of rapidity gaps necessary for the definition of diffractive events~\cite{Kaidalov:2009fp}.  
The apparent breakdown of factorisation in diffractive photo-production also led to approaches where the different components of the cross section, usually the resolved component where the photon assumes a hadronic structure, are rescaled to fit the data~\cite{Klasen:2004qr,Klasen:2008ah}.
A similar rescaling approach was also taken by the authors of~\cite{Zlebcik:2011kq} who analysed differences between \hone and \zeus data as well.
Generally speaking, the resolved components were found to need a suppression by a factor of about 3, however, it remains unclear whether a rescaling of the resolved component only correctly reflects the process of factorization breaking.

Our publication is organised as follows: In Sec.~\ref{Sec:Theo} we will briefly review the theory ingredients for the calculation, in particular the expressions for the pomeron and reggeon fluxes and resulting cross sections, followed by a presentation of our simulation framework and the definition of fiducial phase spaces and experimental observables in Sec.~\ref{Sec:MC}. 
We validate our diffractive DIS simulation in Sec.~\ref{Sec:DDIS}, before turning to a more in-depth discussion of photo-production and the factorization breaking there in Sec.~\ref{Sec:DPHO}.
Predictions for the \EIC for dijet production in both diffractive DIS and diffractive photoproduction in Sec.~\ref{Sec:EIC} provide an outlook to the future.
We summarise our findings in Sec.~\ref{Sec:summary}.

\section{Theory framework for jet production in diffraction}\label{Sec:Theo}

Diffractive events are characterised by the presence of an intact beam proton, or a typically low-mass excitation thereof in the final state.
In~\cite{Collins:1997sr}, factorisation has been shown to hold for these type of processes with the introduction of so-called Diffractive PDFs (DPDF), $f_i^D (x, \mu_F, x_\Pom, t)$.
In this factorised approach, the cross-section for the process $e p \to e X Y$, where $Y$ denotes the elastically scattered proton or its excitation, can be calculated through

\begin{equation}
    \d \sigma_{e p \to e X Y} (x_\Pom, t) = \sum_i f_i^D (x, \mu_F, x_\Pom, t) \otimes \hat{\sigma}_{e i \to e X} (x, \mu_F) \ .
\end{equation}

Due to the colour-disconnection of the final state systems $X$ and $Y$, the events typically show a large rapidity gap (LRG).
The tagging of such events can be experimentally achieved either through dedicated detection of the system $Y$ or by vetoing activity in certain rapidity regions, hence selecting events with LRGs.

Such event topologies are usually attributed to the exchange of Pomeron (and subleadingly, a Reggeon); motivated by this physics model the factorisation picture has been extended~\cite{Ingelman:1984ns}, to separate the DPDF further into products of flux factors and PDFs:

\begin{equation}
    f^D_i \left( x, \mu_F, x_\Pom, t \right) = f_{\Pom / P} \left( x_\Pom, t \right) f_{i / \Pom} \left( x, \mu_F\right)+ n_\Reg f_{\Reg / P} \left( x_\Pom, t \right) f_{i / \Reg} \left( x, \mu_F\right)
\end{equation}

Inspired by Regge theory, the flux factors assume the form

\begin{equation}
    f_{\Pom, \Reg / P} \left( x_\Pom, t \right) = A_{\Pom, \Reg} \frac{e^{B_{\Pom, \Reg}t}}{x_{\Pom, \Reg}^{2 \alpha_{\Pom, \Reg} (t) - 1}}
\end{equation}

with the parameters describing the trajectories $\alpha_{\Pom, \Reg} (t) = \alpha_{\Pom, \Reg} (0) + \alpha_{\Pom, \Reg}^\prime \, t$ fitted to data.
Integrating over $t$ in the interval $[t_{\rm min},\,t_{\rm cut}]$ yields

\begin{equation}
    f_{\Pom, \Reg / P} \left( x_\Pom \right) =
        \frac{A_{\Pom, \Reg} x^{1.\, -2. \alpha_{\Pom, \Reg} (0)} }
             {B_{\Pom, \Reg}-2. \alpha_{\Pom, \Reg}^\prime \log (x)}
        \left( e^{B_{\Pom, \Reg} t_\mathrm{min}} x^{-2. \alpha_{\Pom, \Reg}^\prime t_\mathrm{min}}
             - e^{B_{\Pom, \Reg} t_\mathrm{cut}} x^{-2. \alpha_{\Pom, \Reg}^\prime t_\mathrm{cut}} \right)\,.
\end{equation}

The integration boundaries are given by $t_\mathrm{min} = - \frac{m_p^2 x^2}{1-x}$, with $m_p$ the proton mass, and $t_\mathrm{cut}$ the experimental boundary of the momentum transfer of the proton.
While the PDF for the Pomeron $f_{i / \Pom} \left( x, \mu_F\right)$ has to be fitted to data, the Reggeon PDF is usually approximated by the pion PDF, $f_{i / \Reg} \left( x, \mu_F\right)\approx f_{i / \pi} \left( x, \mu_F\right)$~\cite{Gluck:1991ey}.

Depending on the virtuality of the exchanged photon, events can be further differentiated into the deep-inelastic scattering (DIS) or the photoproduction regime.
In the latter case, the factorisation includes a photon flux and, potentially, a photon PDF.
The overall phase space setup and the implementation of the photoproduction events follows~\cite{Hoeche:2023gme}.

This results in the following factorisation formula for the cross-section for Diffractive DIS

\begin{equation}
    \sigma^\mathrm{(DDIS)} \left( e p \to e X Y \right) =
    \int_{0}^{x_{\Pom{},\mathrm{max}}} \d x_\Pom \int_{t_\mathrm{cut}}^{t_\mathrm{min}} \d t \int_{0}^{1} \d x_i f^D_i \left( x_i, \mu_F, x_\Pom, t \right) \, \hat{\sigma} \left( e i \to e X Y \right)
\end{equation}

and for Diffractive Photoproduction

\begin{align}
    \sigma^\mathrm{(DPHO)} \left( e p \to e X Y \right) =
    &\int_{0}^{1} \d x f^{(e)}_\gamma \left( x \right)  \int_{0}^{x_{\Pom{},\mathrm{max}}} \d x_\Pom \int_{t_\mathrm{cut}}^{t_\mathrm{min}} \d t \nonumber\\
    &\int_{0}^{1} \d x_j f^{(\gamma)}_j \left( x_j, \mu_F \right)
    \d x_i f^D_i \left( x_i, \mu_F, x_\Pom, t \right)
    \ \hat{\sigma} \left( j i \to X Y \right)
\end{align}

where in the latter case $f^{(\gamma)}_j$ is replaced by a Delta distribution for direct photoproduction.
\section{Validation: simulation and data}\label{Sec:MC}

\subsection{Event generation}

Events are generated with a pre-release of \Sherpa~\cite{Gleisberg:2008ta,Sherpa:2019gpd} v3.0.0; the code will be made available in a future release.
The matrix element part of the simulation used \Amegic~ \cite{Krauss:2001iv,Gleisberg:2007md} and \Comix~\cite{Gleisberg:2008fv} for tree-level matrix elements, \OpenLoops~\cite{Cascioli:2011va,Denner:2002ii,Denner:2005nn,Denner:2010tr} for loop matrix elements, and Catani-Seymour dipole subtraction~\cite{Catani:1996vz,Catani:2002hc} automated in~\cite{Gleisberg:2007md} for the treatment of infrared divergences.
The matching to the parton shower~\cite{Schumann:2007mg} is achieved through the \MCatNLO formalism~\cite{Frixione:2002ik} in its implementation presented in~\cite{Hoeche:2011fd}.
For the PDFs we used built-in interfaces to the SAS1M set~\cite{Schuler:1995fk,Schuler:1996fc} for the photon and the H1 2007 Fit B set~\cite{H1:2006zyl} for the pomeron.
For the reggeon and the proton, we used the sets GRVPI0~\cite{Gluck:1991ey} and PDF4LHC21\_40\_pdfas~\cite{PDF4LHCWorkingGroup:2022cjn}, respectively, interfaced through \LHAPDF~\cite{Buckley:2014ana}.
The parameters of pomeron flux were taken from the fit in~\cite{H1:2006zyl}.
We calculated in the 3-flavour scheme in accordance with the PDF and additionally allowed for massive $c$-quarks at NLO and massive $b$-quarks at LO in the final state.
We consistently used the current default value for $\alpha_S = 0.118$ with three-loop running, which is also in-line with the H1 2006 PDF.
The factorisation scale and the renormalisation scale were set to $\mu_F = \mu_R = H_T/2$ for photoproduction events and to $\mu_F^2 = \mu_R^2 = \frac{1}{4} (Q^2 + H_{T, \mathrm{hadr}}^2)$ for DIS events, with $H_{T, \mathrm{hadr}}$ as the scalar sum over the transverse momenta of all hadronic particles, and were varied by factors of 2 in a 7-point scale variation to estimate the uncertainties.
The partonic final states were hadronized with the cluster fragmentation model of \Ahadic~\cite{Chahal:2022rid}, fitted to LEP data.

Diffractive events are simulated with an assumed intact beam proton.
Low-mass excitations have been seen to account for an additional 20\% in cross-section flat in phase space and can therefore be taken into account with an overall scaling of 0.83~\cite{Aaron:2010aa}.
Diffractive photoproduction is composed of two components, the direct and the resolved photon contributions, with the latter simulated through a photon PDF.
However appealing this picture of combining two PDFs is -- one for the photon, one for the pomeron or reggeon -- the assumed factorisation underpinning it is expected to break down~\cite{Collins:1997sr} as a consequence of additional soft interactions between the photon and the proton beam.
To account for the suppression of these events, we generalized the multiple-interactions modelling in \Sherpa to allow for this kind of interactions.
Naturally, this argument can only be applied to the resolved component and therefore it has been conjectured that factorisation might still hold for the direct component~\cite{Kaidalov:2003xf}.
We will study this ansatz in Sec. \ref{Sec:FactorisationBreaking}.

\subsection{Experimental observables and datasets}\label{Sec:Exp}

Our implementation of diffractive events in DIS and photoproduction was validated with data from the \hone~\cite{H1:2015okx,H1:2007jtx} and \zeus~\cite{ZEUS:2007uvk} collaborations.

\paragraph{H1, JHEP05 (2015) 056}
The data from the \hone collaboration in~\cite{H1:2015okx} detected the out-going proton in the Very Forward Proton Spectrometer (VFPS) and measured di-jet production in both the DIS, $4 \UGeV^2 < Q^2 < 80 \UGeV^2$, and photoproduction, $Q^2 < 2 \UGeV^2$, regime. To describe the kinematics of the diffractive exchange, the variable $x_\Pom$ was defined as

\begin{equation}
    x_\Pom = 1 - \frac{E_p^\prime}{E_p}
\end{equation}

with $E_p^{(\prime)}$ the energy of the incoming (outgoing) proton; the acceptance of the VFPS yielded a range of $0.010 < x_\Pom < 0.024$.
The partonic momentum fractions with respect to the diffractive exchange and, in the case of photoproduction, with respect to the electron and the photon have been defined as

\begin{equation}
    z_\Pom = \frac{Q^2 + M_{12}^2}{Q^2 + M_X^2}
\end{equation}

for diffractive DIS events and

\begin{align} 
    z_\Pom &= \frac{\sum_{j = 1, 2} \left( E + p_z\right)_j}
                   {\sum_{i \in X} \left( E + p_z\right)_i}\;,\;\;\;
    y = \frac{\sum_{i \in X} \left( E - p_z\right)_i}{2 E_e} \;,\;\mbox{\rm and}\;\;
    x_\gamma = \frac{\sum_{j = 1, 2} \left( E - p_z\right)_j}
                    {\sum_{i \in X} \left( E - p_z\right)_i}
\end{align}

for diffractive photoproduction events, where $M_{12}$ denotes the di-jet invariant mass, $M_X$ the invariant mass of $X$ defined below; the index $j$ runs over the leading jets and the index $i$ over the final states in $X$.
The momentum transfer $t$ from the proton was required to be $\left| t \right| < 0.6 \UGeV^2$. The leading and subleading jets were clustered with the $k_T$-algorithm with $R = 1$ in the photon-proton rest frame and were required to have transverse energies of $E_T^* > 5.5 \UGeV$ and $4.0 \UGeV$\footnote[2]{By the superscript '$^*$' we denote quantities measured in the photon-proton rest frame.}, respectively, and lie within the pseudorapidity range of $-1 < \eta < 2.5$ in the laboratory frame. The inelasticity $y$ was required to be within $0.2 < y < 0.7$. The invariant mass of the system $X$ was calculated as

\begin{equation}\label{eq:h1-mx}
    M_X^2 = \left( \sum_{i \in X} E_i \right)^2 - \left( \sum_{i \in X} \vec{p}_i \right)^2 \ .
\end{equation}

\paragraph{H1, EPJC51 (2007) 549}
A similar measurement was undertaken in~\cite{H1:2007jtx}, where the mass of the system $Y$ was restricted to $M_Y^2 < 1.6 \UGeV^2$. Similarly to before, di-jets were measured in the DIS, $4 \UGeV^2 < Q^2 < 80 \UGeV^2$, and photoproduction, $Q^2 < 0.01 \UGeV^2$, regimes and jets were clustered as before, demanding $E_T^* > 5 (4) \UGeV$ for the (sub)leading jet within pseudorapditiy $-1 < \eta^\mathrm{lab} < 2$. The photon-proton c.m.s.-energy $W$ was restricted to $165 \UGeV < W < 242 \UGeV$ and the proton momentum transfer to $\left| t \right| < 1 \UGeV^2$. The kinematical variables were defined as

\begin{align}
    x_\Pom = \frac{Q^2 + M_X^2}{Q^2 + W^2} \;,\;\mbox{\rm and}\;\;
    z_\Pom = \frac{Q^2 + M_{12}^2}{Q^2 + M_X^2} \ ,
\end{align}

with $M_{12}$ the dijet mass, for diffractive DIS events and

\begin{align}
    x_\Pom = \frac{\sum_{i \in X} \left( E + p_z\right)_i}
                   {2 E_p} \;,\;\;\;
    z_\Pom = \frac{\sum_{j = 1, 2} \left( E + p_z\right)_j}{2 x_\Pom E_p}\;,\;\;\;
    y = 1 - \frac{E_e^\prime}{E_e} \;,\;\mbox{\rm and}\;\;
    x_\gamma = \frac{\sum_{j = 1, 2} \left( E - p_z\right)_j}{2 y E_e}\nonumber\\
\end{align}

for diffractive photoproduction events, where $E_e^{(\prime)}$ denotes the incoming (outgoing) lepton's energy; $M_X$ was defined as in the previous analysis, c.f. Eq.~\ref{eq:h1-mx}.

\paragraph{ZEUS, EPJC55 (2008) 177}
The \ZEUS collaboration measured di-jet production in diffractive photoproduction~\cite{ZEUS:2007uvk} with photon virtualities of $Q^2 < 1 \UGeV^2$ and an inelasticity of $0.20 < y < 0.85$. The out-going proton was not detected, instead diffractive events were selected by requiring LRGs. To correct for proton-dissociative events, a fraction of $16 \%$ had been subtracted from the total cross-section. The following variables were defined:

\begin{align}
    x_\Pom = \frac{\sum_h \left( E + p_z\right)_h}{2 E_p} \;,\;\;\;
    z_\Pom = \frac{\sum_j E_T^{(j)} \mathrm{e}^{\eta^{(j)}}}{2 x_\Pom E_p}\;,\;\;\;
    y &= \frac{\sum_h \left( E - p_z\right)_h}{2 E_e} \;,\;\;\;
    x_\gamma = \frac{\sum_j E_T^{(j)} \mathrm{e}^{-\eta^{(j)}}}{2 y E_e} \;,
\end{align}
and
\begin{align}
    M_X^2 = \sum_h \left( E - p_z\right)_h \left( E + p_z\right)_h\;,
\end{align}

where the index $h$ runs over reconstructed Energy Flow Objects in the main detectors and it was required $x_\Pom < 0.025$. Jets were clustered with the $k_T$ algorithm using $R = 1$ in the laboratory frame, with cuts of $E_T > 7.5 (6.5) \UGeV$ for the (sub)leading jet and within pseudorapidity $-1.5 < \eta < 1.5$.

\section{Diffractive DIS}\label{Sec:DDIS}

\begin{figure}[!ht]
    \centering
    \begin{tabular}{ccc}
        \includegraphics[width=.3\linewidth]{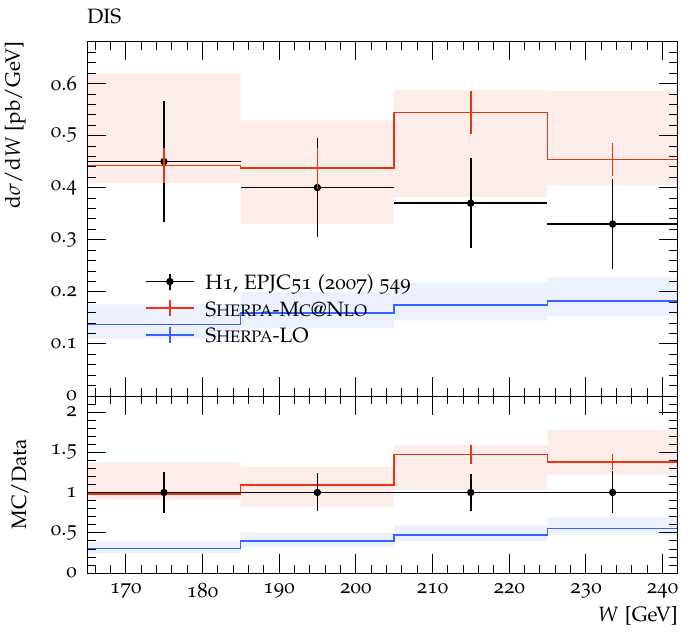} &
        \includegraphics[width=.3\linewidth]{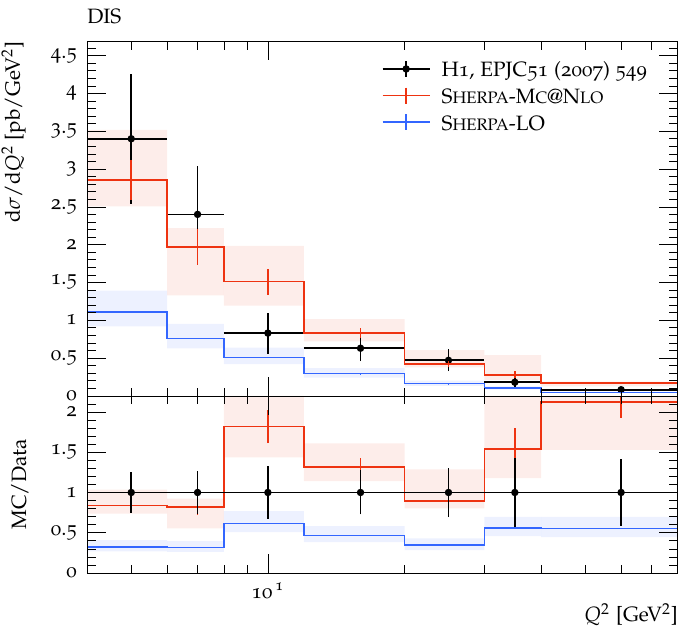} &
        \includegraphics[width=.3\linewidth]{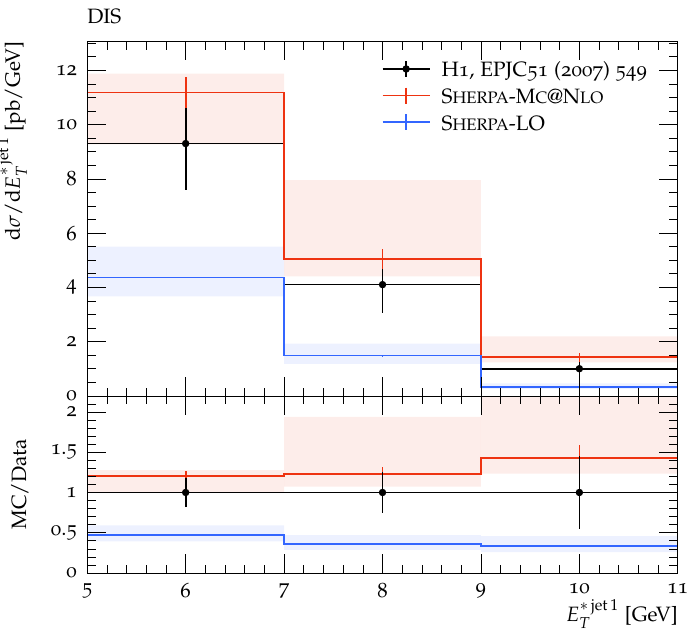} \\
        \includegraphics[width=.3\linewidth]{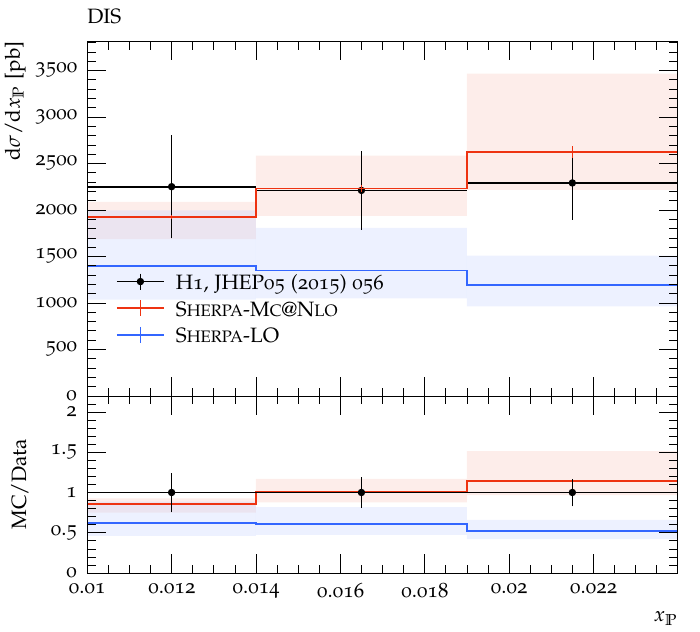} &
        \includegraphics[width=.3\linewidth]{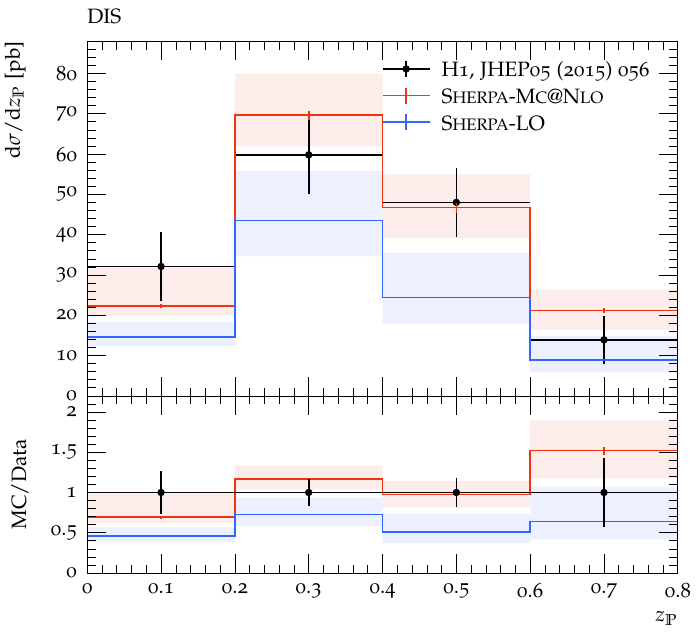} &
        \includegraphics[width=.3\linewidth]{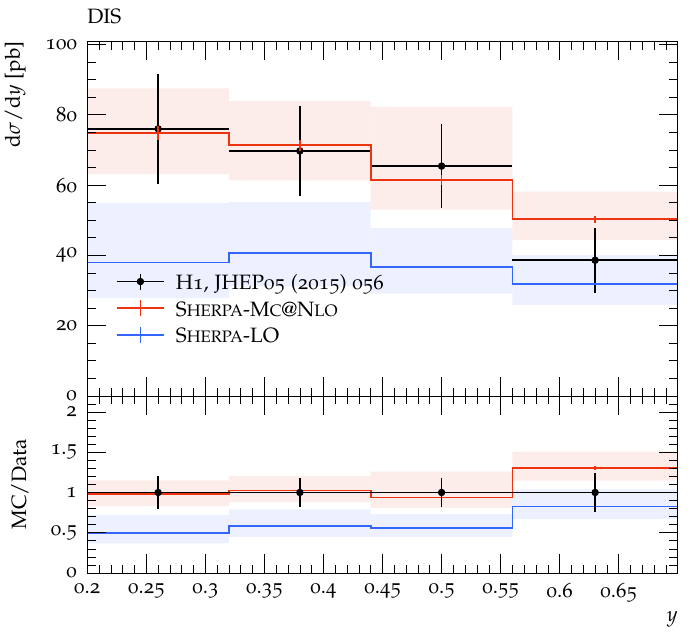}\\
        \includegraphics[width=.3\linewidth]{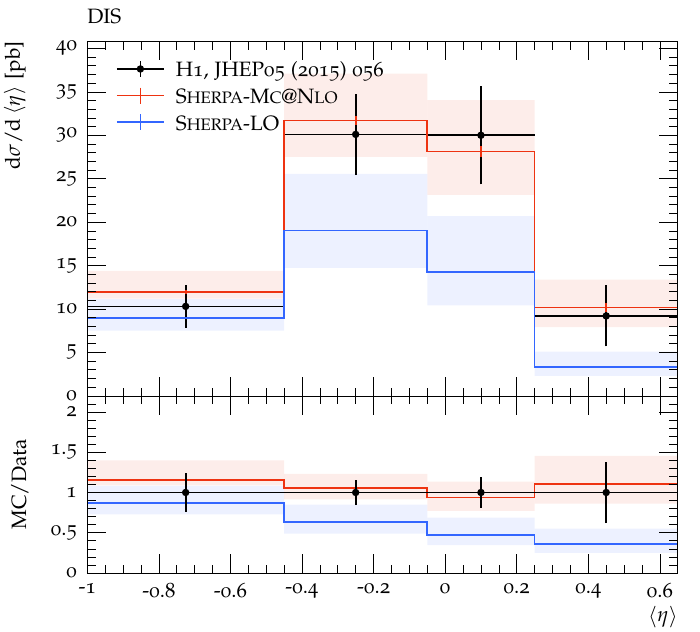} &
        \includegraphics[width=.3\linewidth]{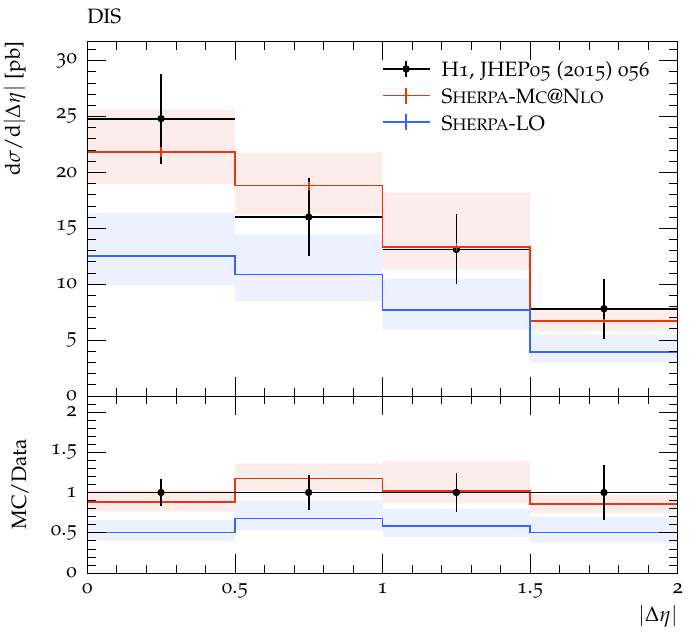} &
        \includegraphics[width=.3\linewidth]{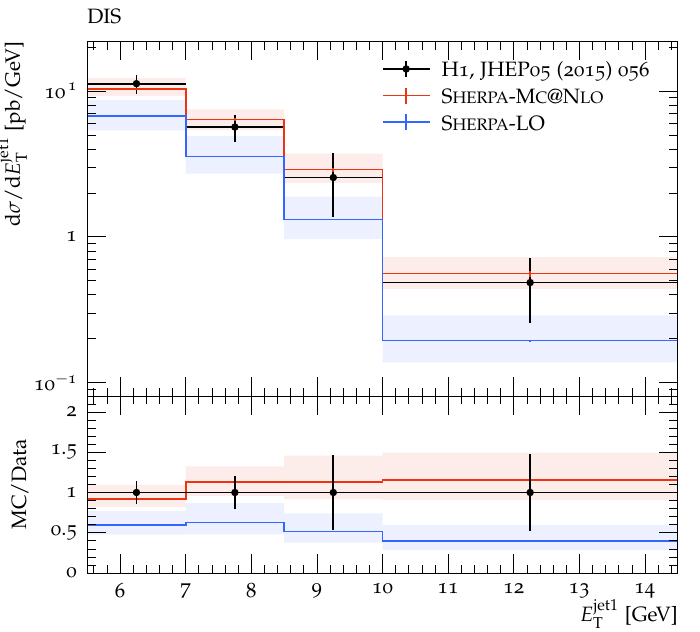}
    \end{tabular}
    \caption{Differential DDIS cross-sections with respect to the photon-proton centre-of-mass energy $W$, the photon virtuality $Q^2$, and the leading jet transverse momentum $E_T^{*\rm{jet 1}}$ (upper row, left to right, data from~\protect\cite{H1:2007jtx}), to momentum ratios $x_\Pom$, $z_\Pom$ and inelasticity $y$ (second row, left to right, data from~\protect\cite{H1:2015okx}),  to the average pseudorapidity $\langle\eta\rangle$ and the pseudorapidity difference $\Delta\eta$ of the jets as well as the transverse momentum of the leading jet $E_T^{\rm{jet1}}$ (third row, left to right, data from~\protect\cite{H1:2015okx}).}
    \label{fig:nlo-lo-ddis-h1-1}
\end{figure}
Turning first to the Diffractive DIS measurements, we compare the \Sherpa results, obtained at \MCatNLO accuracy with dijet production data from two publications by the \hone collaboration~\cite{H1:2007jtx,H1:2015okx}.

In the upper two rows of Fig.~\ref{fig:nlo-lo-ddis-h1-1} we focus on more inclusive observables that describe the overall kinematics of the events.
They include very general observables such as the centre-of-mass energy of the photon-proton system $W$, the photon virtuality $Q^2$ or the transverse momentum of the leading jet, $E_T^{*{\rm jet1}}$, all taken from~\cite{H1:2007jtx}, and extend to observables that expose more of the underlying parton-level dynamics such as the momentum fractions $x_\Pom$ and $z_\Pom$ and the inelasticity $y$, with data from~\cite{H1:2015okx}.
We consistently observe excellent agreement of the simulation at \MCatNLO accuracy with data, with $K$-factors ranging from about 1.5 up to 2-3 depending on the observable and the associated phase space region.
With the exception of the $W$ and $y$ distributions, the $K$-factors are relatively flat and do not overly change the shape of the distributions.
This confirms other findings which established good agreement of data and fixed-order calculations at NLO~\cite{Britzger:2018zvv} and the corresponding $K$-factors.
The increasing size of $K$-factors is readily understood and associated with the additional phase space made available due to the asymmetric cuts and additional parton-level channels with the largest enhancement of the cross-section seen in the forward region, as expected.
\clearpage

\section{Diffractive photoproduction}\label{Sec:DPHO}

\subsection{Diffractive photoproduction at \MCatNLO accuracy}
\begin{figure}[!ht]
    \centering
    \begin{tabular}{cc}
        \includegraphics[width=.35\linewidth]{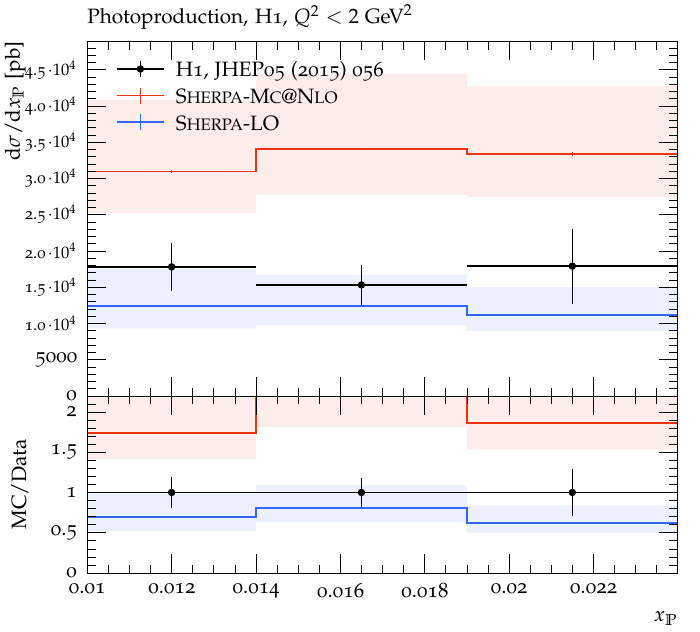} &
        \includegraphics[width=.35\linewidth]{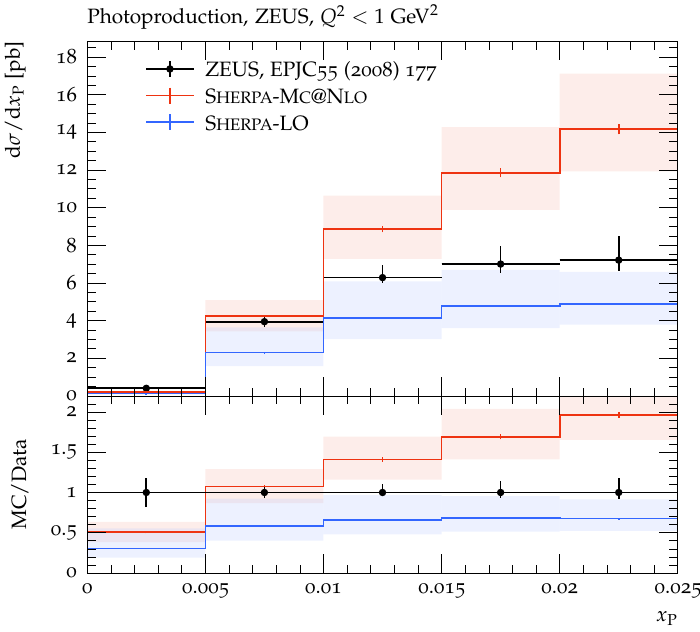} \\
        \includegraphics[width=.35\linewidth]{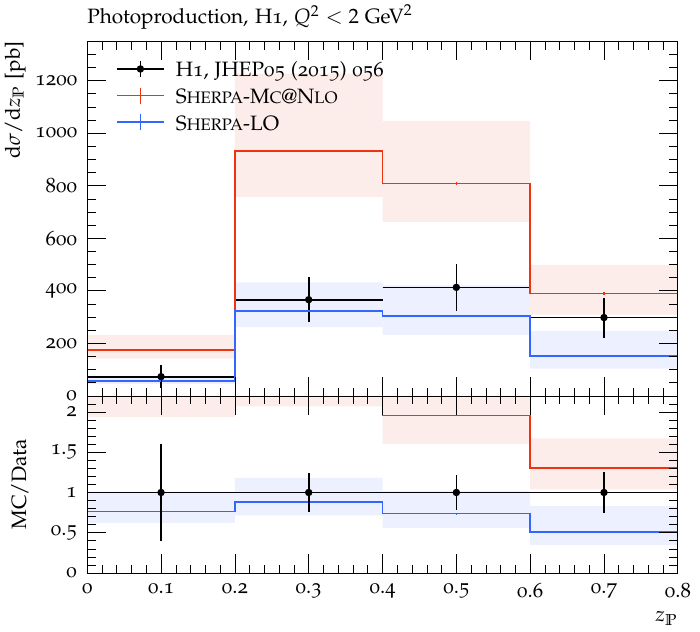} &
        \includegraphics[width=.35\linewidth]{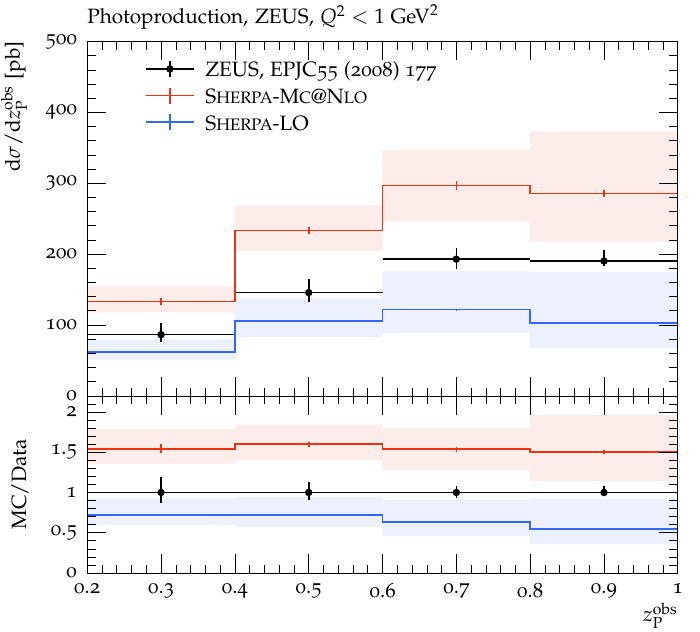} \\
        \includegraphics[width=.35\linewidth]{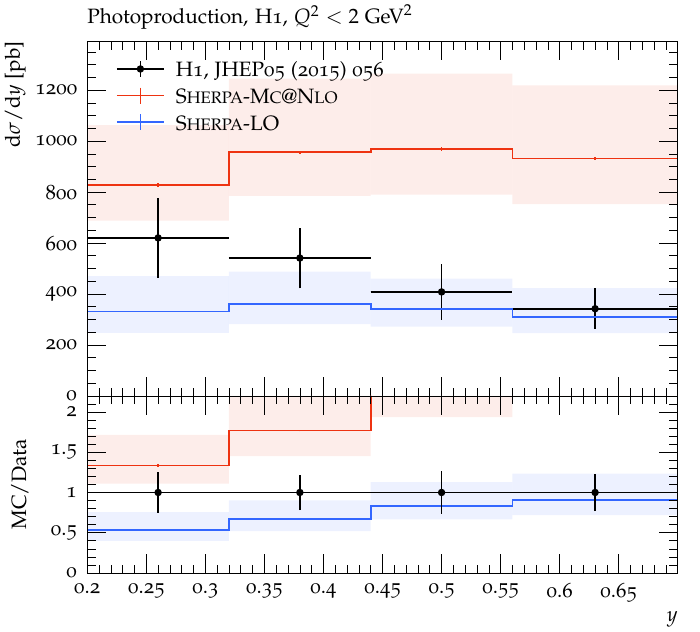} &
        \includegraphics[width=.35\linewidth]{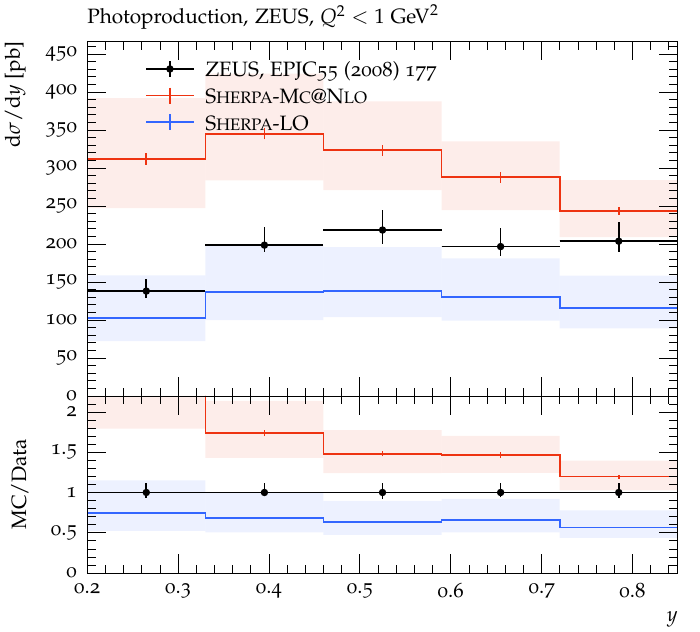} \\
        \includegraphics[width=.35\linewidth]{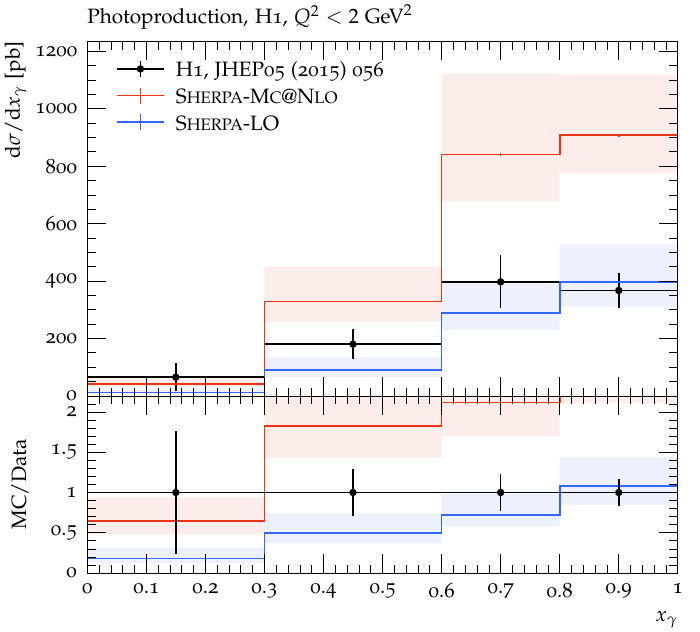} &
        \includegraphics[width=.35\linewidth]{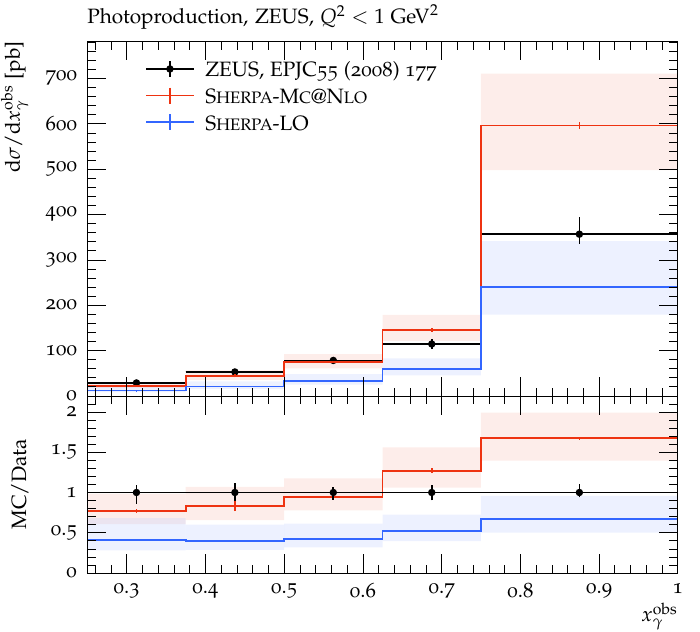}
    \end{tabular}
    \caption{Differential diffractive photoproduction cross-sections with respect to momentum ratios $x_\Pom$, $z_\Pom$, $y$ and $x_\gamma$ (top to bottom), obtained by \hone (left column, data from~\protect\cite{H1:2015okx}) and \zeus (right column, data from~\protect\cite{ZEUS:2007uvk}).}
    \label{fig:nlo-lo-dpho-x-h1-zeus}
\end{figure}
Turning now to the description of diffractive photoproduction of dijets, we observe that the convincing agreement of simulation and data does not hold true anymore, supporting statements about the possible breakdown of factorization in such processes~\cite{Collins:1997sr}.
To highlight this, let us first take a look at the momentum ratios $x_\Pom$, $z_\Pom$, $y$ and $x_\gamma$.
They have been analysed by both the \hone and the \zeus collaboration in~\cite{H1:2015okx} and~\cite{ZEUS:2007uvk}, respectively, and we display the results in the left and right column of Fig.~\ref{fig:nlo-lo-dpho-x-h1-zeus}.
As already seen in previous NLO calculations~\cite{Klasen:2004qr}, the calculation of the cross-section in our \MCatNLO samples severely overestimates the data in diffractive photoproduction, by a factor of up to 2-3, while the LO predictions are in somewhat better agreement overall.
In addition, we observe a sharp increase, amounting to a visible shape distortion, of the \MCatNLO simulation with respect to the experiment, particularly at large values of $x_\gamma\stackrel{>}{\sim}0.75$, a regime usually associated with "direct" photoproduction, i.e.\ the photon acting as a point-like particle.
By far and large, however, the \MCatNLO simulation describes data reasonably well in the "resolved" photoproduction regime of small $x_\gamma\stackrel{<}{\sim}0.5$.

We note that differences in the shape and normalisation of between the two different sets of data - and therefore between the two simulations - stem from the different phase spaces populated by the \hone and \zeus analyses, mainly related to the difference in the range of $Q^2$ and the definition of the diffractively scattered proton.

\begin{figure}[h!]
    \centering
    \begin{tabular}{cc}
        \includegraphics[width=.4\linewidth]{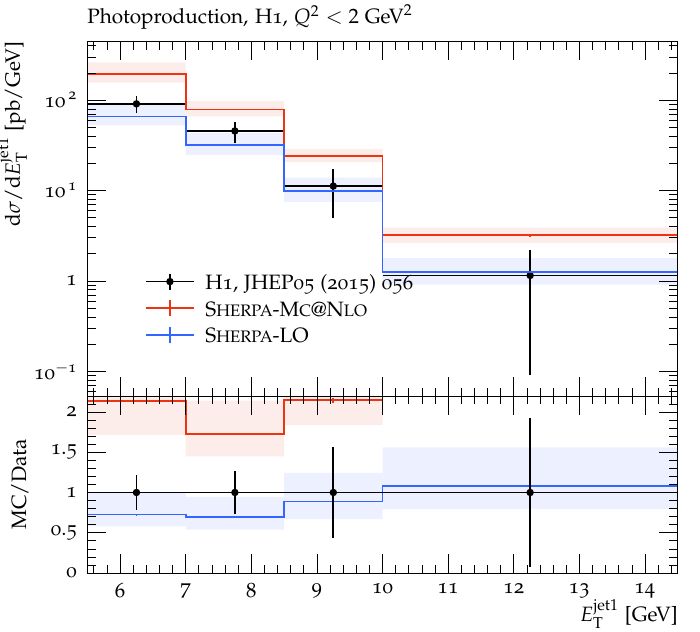} &
        \includegraphics[width=.4\linewidth]{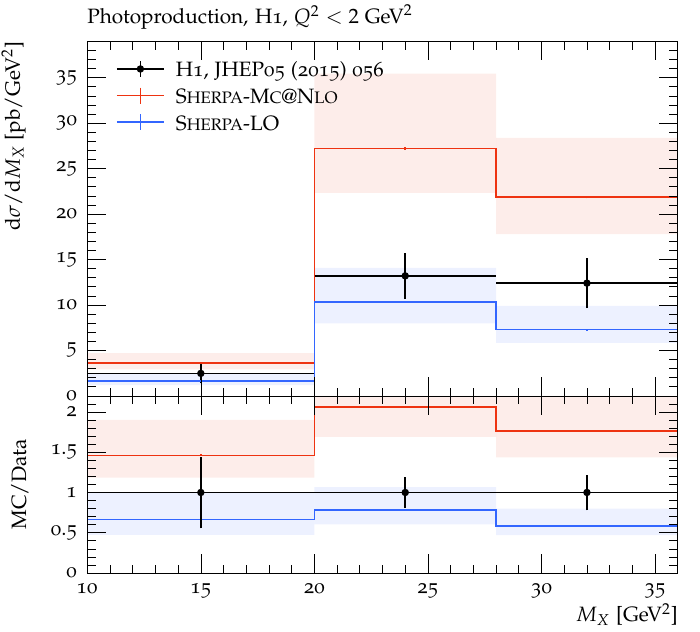} \\
        \includegraphics[width=.4\linewidth]{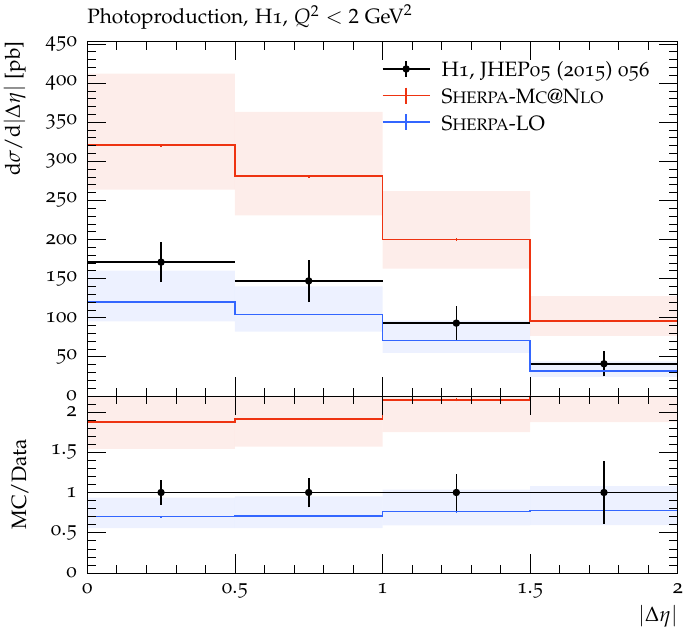} &
        \includegraphics[width=.4\linewidth]{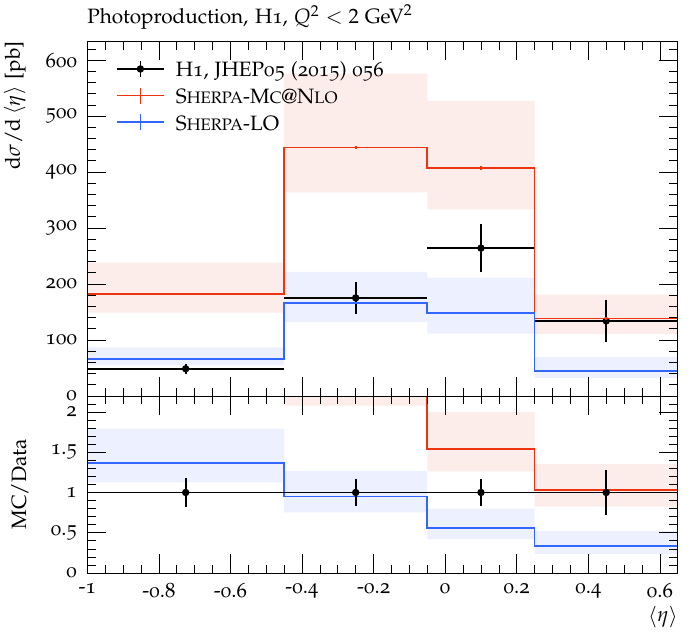}
    \end{tabular}
    \caption{Differential dijet Diffractive Photoproduction cross-sections with respect to leading jet transverse energy $E_T^\text{jet1}$, invariant mass $M_X$, difference in dijet-pseudorapidity $|\Delta\eta|$ and average dijet-pseudorapditiy $<\eta>$. Results of the \protect\Sherpa simulation with \MCatNLO accuracy are compared with results at LO and with data from \protect\hone~\protect\cite{H1:2015okx}. }
    \label{fig:nlo-lo-dpho-dijet-h1}
\end{figure}

Our findings so far of NLO predictions overshooting data by large factors, are in agreement also with an analysis of final state observables, such as the leading jet transverse energy, $E_T^{\rm jet1}$, the diffractive invariant mass $M_X$, and the average pseudorapidity of jets $\langle\eta\rangle$ and the pseudorapidity difference on the two leading jets $|\Delta\eta|$, which we show in Fig.~\ref{fig:nlo-lo-dpho-dijet-h1}.
Again the \MCatNLO predictions have a $K$-factor of about 2 with respect to their leading-order counterparts, and they exceed data (taken from \protect\hone~\protect\cite{H1:2015okx}), again, by a factor of about 2.
We observe that the shapes of the distributions do not agree between data and theory, especially in the pseudorapditiy distributions and the momentum fractions.

To compensate for the massive difference in measured and calculated cross sections, \hone applied a global factor of about 1/2 to the calculation~\cite{H1:2007jtx,H1:2010xdi,H1:2015okx}, improving its agreement with the data.
This rescaling does not improve the theory agreement with the \zeus data, where the calculation even slightly undershoots the data for low $x_\gamma$; therefore a global scaling can not adequately be used to describe the factorisation breaking.
We will turn to this problem in more detail in subsubsection~\ref{Sec:FactorisationBreaking} by further examining the interplay seen in this observable.

\subsection{Reggeon contribution}
\begin{figure}[h!]
    \centering
    \begin{tabular}{ccc}
        \includegraphics[width=.3\linewidth]{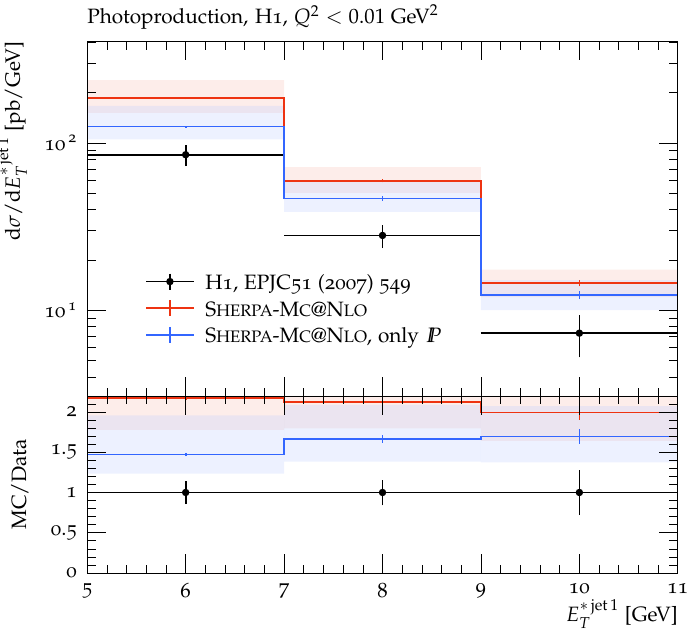} &
        \includegraphics[width=.3\linewidth]{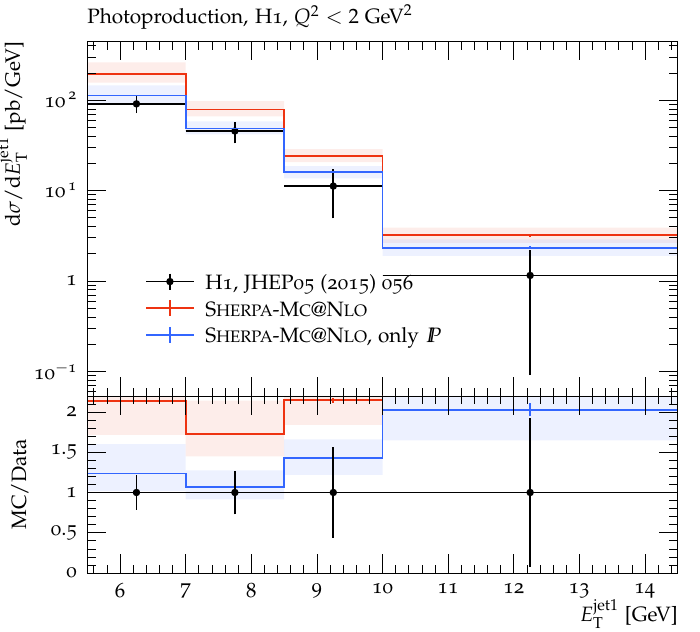} &
        \includegraphics[width=.3\linewidth]{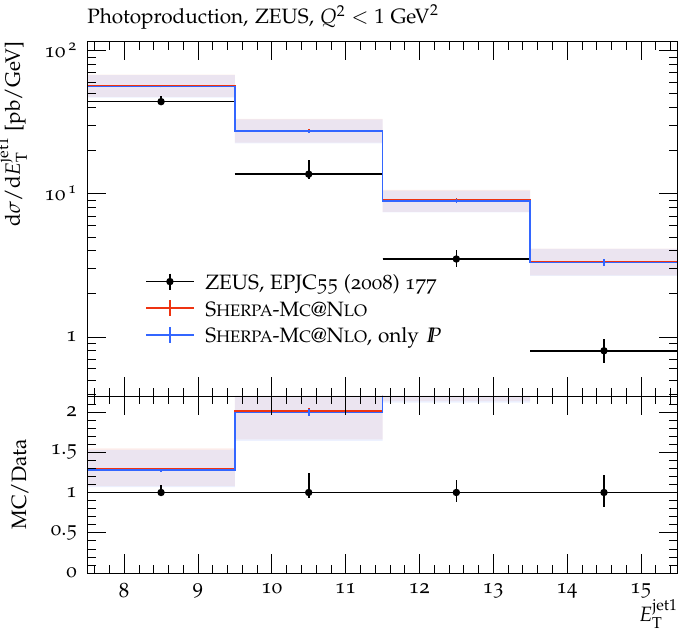} \\
        \includegraphics[width=.3\linewidth]{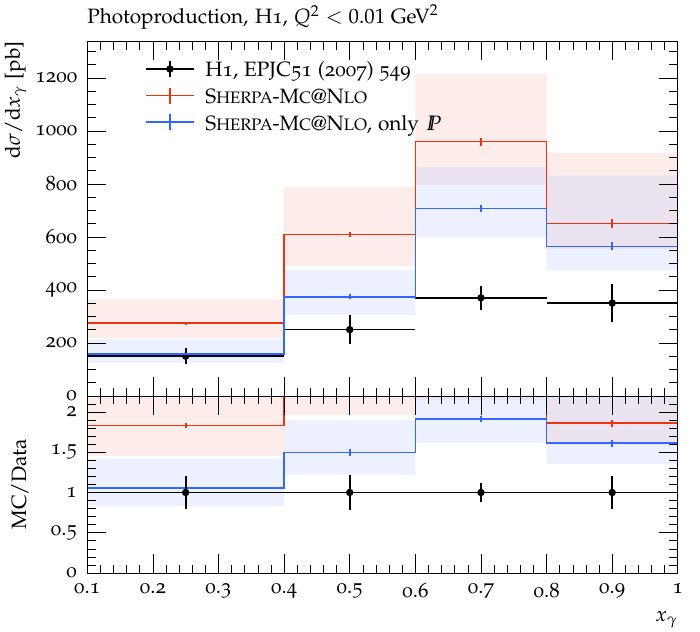} &
        \includegraphics[width=.3\linewidth]{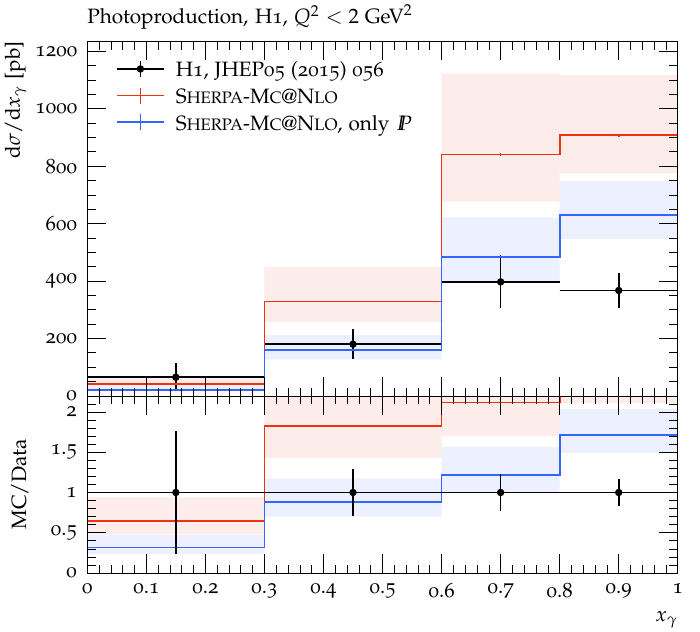} &
        \includegraphics[width=.3\linewidth]{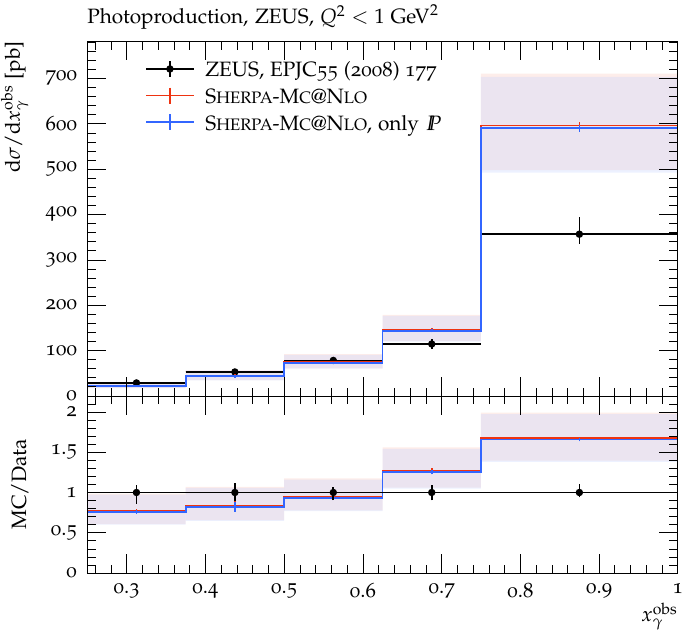}
    \end{tabular}
    \caption{Distributions of leading jet transverse energy $E_T^\text{jet1}$ (top row) and $x_\gamma$ (bottom row) with and without Reggeon contribution, compared to data by \protect\hone from~\protect\cite{H1:2007jtx} (left) and~\protect\cite{H1:2015okx} (middle) and from \protect\zeus~\protect\cite{ZEUS:2007uvk} (right) in their respective definitions of fiducial phase space. }
    \label{fig:reggeon-contrib-et}
\end{figure}
To further elucidate the components of the cross-section, we show in Fig.~\ref{fig:reggeon-contrib-et} the leading jet transverse energy, comparing the calculation with and without the Reggeon contribution, in three different definitions of phase space.
Depending on it we find sizeable positive contribution for the full range of the observable.

\clearpage

\subsection{Modelling factorisation breaking in diffractive photoproduction}\label{Sec:FactorisationBreaking}

The overshoot of the cross-section hints at the breakdown of the factorisation and different models have been brought forward to explain the discrepancies, which we will review in the following.
The variable $x_\gamma$ has been used by the experiments as a discriminator between the direct and resolved photon components and our calculation confirms that it indeed works well to discern the different contribution.
Generally, a small value of $x_\gamma$ will correspond to a dominant contribution by the resolved photon and its PDF, while values close to unity will mostly stem from direct contributions, i.e.\ the photon acting as a point-like particle.
In this section we use this observable the handle it provides on the dynamics of factorisation breaking and its interplay with the direct and/or resolved components.

\begin{figure}[htpb]
    \centering
    \begin{tabular}{ccc}
        \includegraphics[width=.3\linewidth]{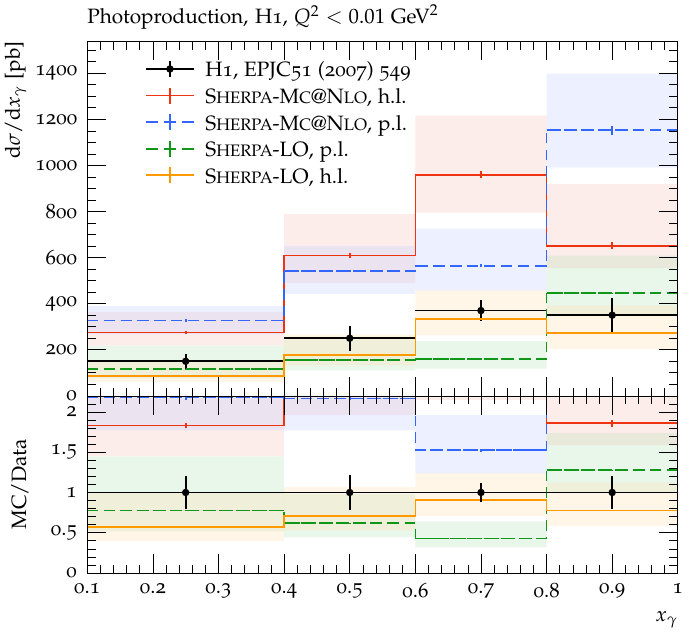} &
        \includegraphics[width=.3\linewidth]{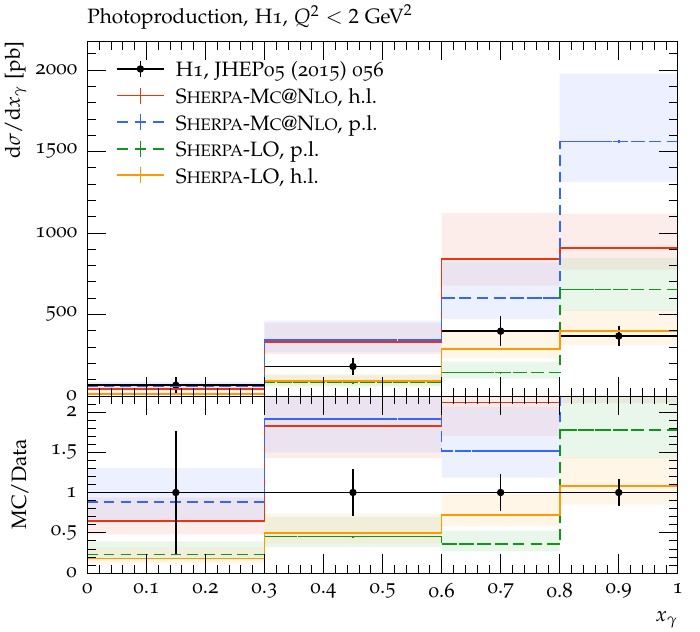} &
        \includegraphics[width=.3\linewidth]{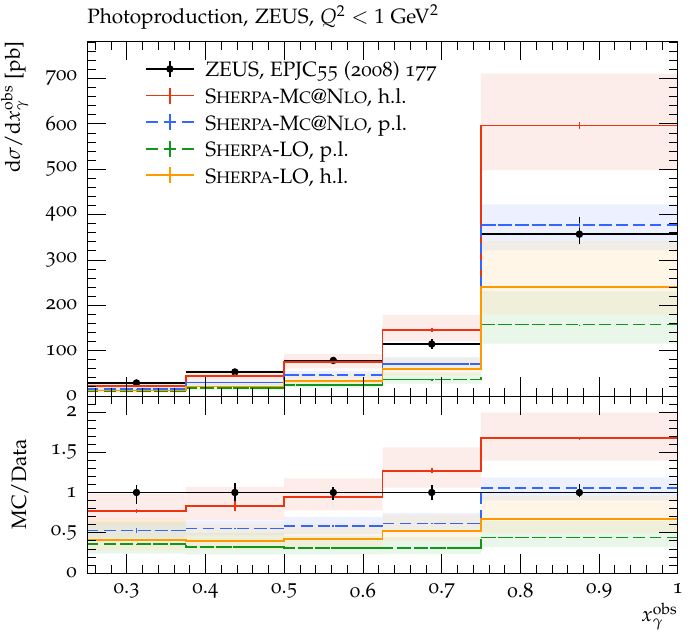}
    \end{tabular}
    \caption{Differential distribution of $x_\gamma$ in diffractive photoproduction. Results of the \protect\Sherpa simulation with \MCatNLO accuracy are compared with results at LO and with data by \protect\hone from~\protect\cite{H1:2007jtx} (left) and~\protect\cite{H1:2015okx} (middle) and \protect\zeus~\protect\cite{ZEUS:2007uvk} (right). }
    \label{fig:pl-hl-dpho-xgamma-h1}
\end{figure}

In~\cite{Kaidalov:2009fp} it was argued that the factorization breaking is a consequence of hadronisation, bin migration and NLO effects, which we exemplify in the distribution of $x_\gamma$ in Fig.~\ref{fig:pl-hl-dpho-xgamma-h1}.
While hadronisation certainly plays a big role in the bin migration between the two largest $x_\gamma$ bins and the total cross-section is decreased, the overall overestimation of the total cross-section is still present after taking these effects into account.
In fact, hadronisation is essential for the reconstruction of $x_\gamma$ in the \zeus analysis, as a significant number of events in the parton-level simulation end up in the the region $x_\gamma > 1$.
It is also somewhat amusing to note that hadronisation effects tend to reduce the overshoot of the simulation in the highest $x_\gamma$ bins for the \hone analyses, while they tend to actually create it for the \zeus analysis.

\begin{figure}[htpb]
    \centering
    \begin{tabular}{ccc}
        \includegraphics[width=.3\linewidth]{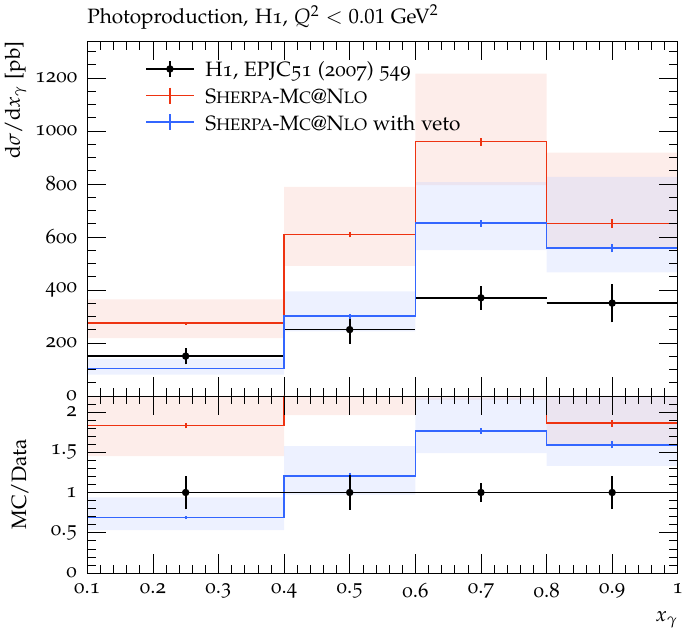} &
        \includegraphics[width=.3\linewidth]{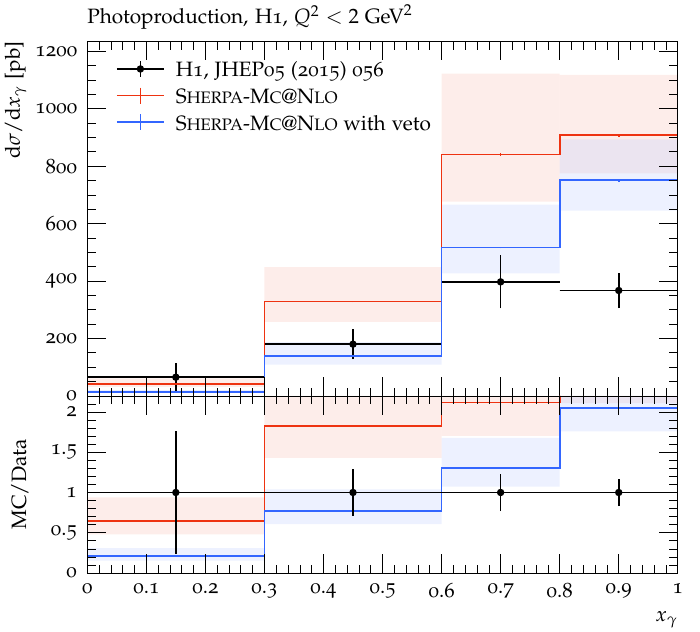} &
        \includegraphics[width=.3\linewidth]{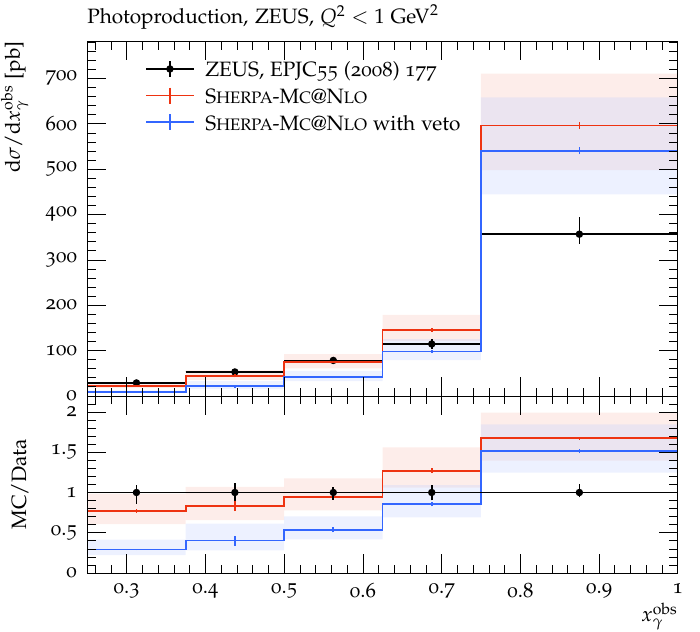}
    \end{tabular}
    \caption{Differential diffractive photoproduction cross-sections with respect to momentum ratio $x_\gamma$ as measured by \protect\hone in~\protect\cite{H1:2007jtx} (left) and~\protect\cite{H1:2015okx} (middle) and $x_\gamma^\mathrm{obs}$ as measured by \protect\zeus~\protect\cite{ZEUS:2007uvk} (right), compared to results of the \protect\Sherpa simulation at \MCatNLO accuracy with and without a veto on $\gamma p$ interactions. }
    \label{fig:bbr-dpho-x}
\end{figure}

As mentioned previously, it has furthermore been argued that the decrease of the cross-section is due to soft interactions between the resolved photon and the proton~\cite{Collins:1997sr}.
We implemented a simplistic generalized multiple-parton interactions modelling in \Sherpa to veto events, which have an additional scattering between the photon and the proton and would hereby destroy the rapidity gap.
In Fig.~\ref{fig:bbr-dpho-x} we show the effect of this rejection.
Naturally this mechanism only applies to the resolved component and affects only those regions of the phase space where the resolved component dominates.
We also recall that the resolved component can be further decomposed into the point- and hadron-like component, where the difference is that the backwards evolution would collapse the former to a photon and the latter to a meson-like state.
Furthermore, it has been pointed out in~\cite{Kaidalov:2009fp} that this so-called anomalous component, i.e.\ splittings of $\gamma \to q \bar{q}$, would not exhibit further interactions with the proton.
This would lead to a smaller suppression in the resolved component, depending on the size of these splittings in the radiation off these quarks\footnote{
These splittings are currently not included in the simulation in \Sherpa and, as indicated, we leave the study of this effect to future work.}.
We therefore expect that the generalized MPI modelling would only apply to meson-like states, whereas the point-like contribution would see a suppression mechanism similar to the direct component.
A comprehensive study of the suppression mechanism will have to disentangle these two components; while the hadron-like states will undergo the MPI-based breaking of the factorisation, the point-like state will have to interpolate between the direct and the hadron-like regimes.
The implementation of this modelling and the details of the suppression of the point-like resolved component are left for future work.
However, while our naive model depends on some assumptions of the impact parameter and other unconstrained parameters and some additional simplifications, which certainly deserve further investigation, it does not appear as if these effects alone can accommodate the observed large discrepancy of simulation and data.

The authors of~\cite{Zlebcik:2011kq} conducted a study which found that, even though there is a slight dependence, the different phase space cuts are not the cause of the discrepancy in the suppression between the \hone and \zeus measurements.

\begin{table}[h!]
    \begin{center}
        \begin{tabular}{|l||c|c|c|}
        \hline
         & \hone, EPJC51 (2007) 549~\cite{H1:2007jtx} & \hone, JHEP05 (2015) 056~\cite{H1:2015okx} & \zeus, EPJC55 (2008) 177~\cite{ZEUS:2007uvk} \\
        \hline\hline
        $R_\mathrm{res}$ & $0.4 \pm 0.1$ & $0.6 \pm 0.3$ & $1.3 \pm 0.1$ \\ \hline
        $R_\mathrm{dir}$ & $0.4 \pm 0.1$ & $0.3 \pm 0.2$ & $0.5 \pm 0.1$ \\
        \hline
        \end{tabular}
    \parbox{0.8\textwidth}{\caption{
        Scaling factors for the direct, $R_\mathrm{dir}$, and resolved, $R_\mathrm{res}$, component in diffractive photoproduction for the respective experimental data.  }
        \label{tab:fitting-factors}
    }
    \end{center}
\end{table}
\begin{figure}[h!]
    \centering
    \begin{tabular}{ccc}
        \includegraphics[width=.3\linewidth]{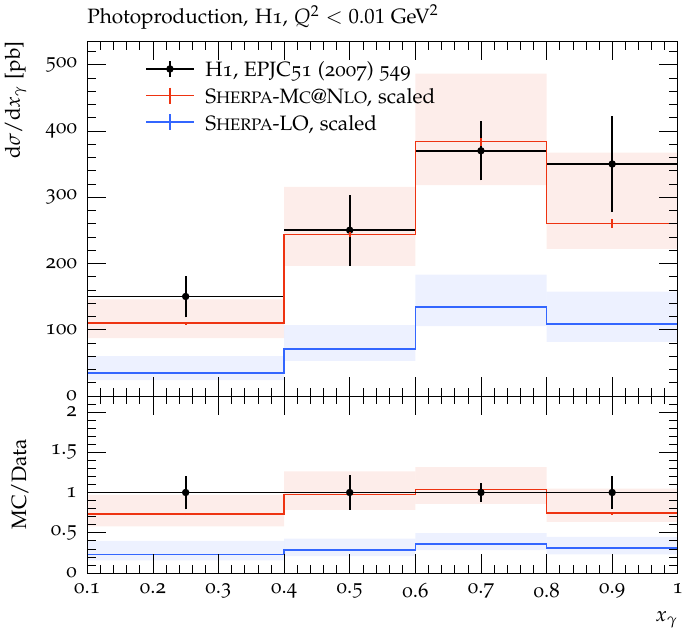} &
        \includegraphics[width=.3\linewidth]{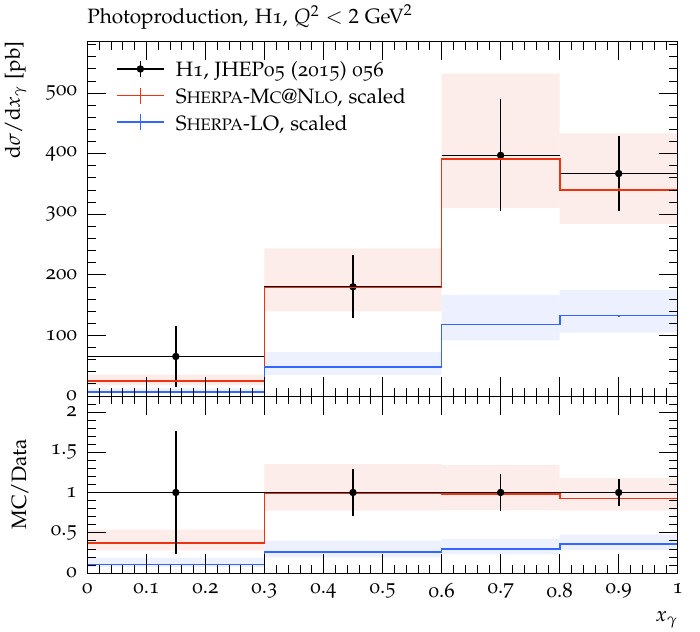} &
        \includegraphics[width=.3\linewidth]{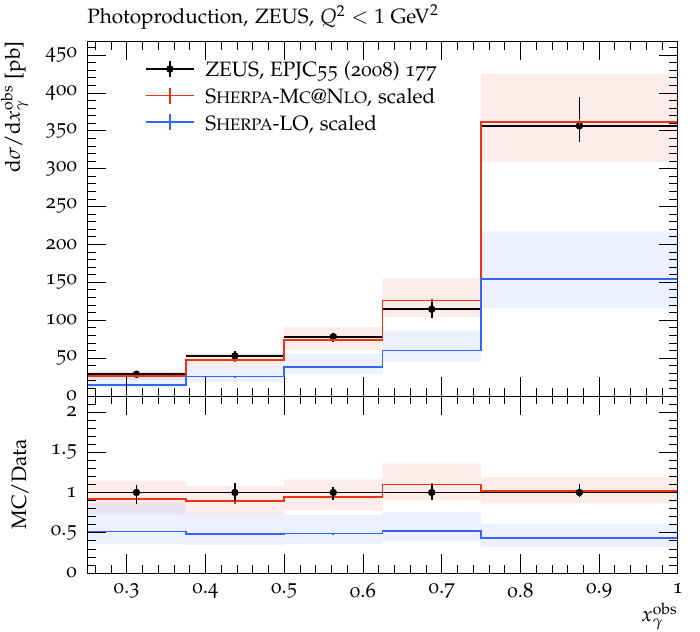}
    \end{tabular}
    \caption{Differential diffractive photoproduction cross-sections with respect to momentum ratio $x_\gamma$ as measured by \protect\hone in~\protect\cite{H1:2007jtx} (left) and~\protect\cite{H1:2015okx} (middle) and $x_\gamma^\mathrm{obs}$ as measured by \protect\zeus~\protect\cite{ZEUS:2007uvk} (right), compared to results of the \protect\Sherpa simulation at \MCatNLO accuracy with the direct and resolved component scaled separately. }
    \label{fig:dpho-fitting}
\end{figure}
Turning back to the distributions of $x_\gamma$, we point out that, while the data from \zeus and \hone do not agree with each other on the overall size of the necessary suppression, they both show an overshoot for large $x_\gamma$ values.
This hints at factorisation breaking happening in the direct component too.
In fact, the distinction between direct and resolved components of photoproduction can not be maintained at NLO as the real correction to "direct" photoproduction, $\gamma j \to jjj$, can not be distinguished from contributions of photon splitting combined with a two-jet matrix element, $\left( \gamma \to q \bar{q} \right)_\mathrm{PDF} \otimes \left( qj \to jj \right)_\mathrm{ME}$, where $j$ denotes any parton.
This led us to revisit the logic outlined in~\cite{H1:2007jtx}, to further elucidate the impact of factorisation breaking on the different photonic components.
We fitted two prefactors, one each for direct and resolved components in the simulation, to the data to quantify the effect of the suppression in the two components separately, with results shown in Tab.~\ref{tab:fitting-factors}.
While the scaling of the resolved component does vary among the different measurements, the direct component seems to support a somewhat universal suppression by a factor of 0.5.
The data covered different cuts on the photon virtuality, hence the suppression seems to be independent of the kinematics at the electron-photon vertex.

In Fig.~\ref{fig:dpho-fitting} we exhibit the results of the fit; we also did not observe any significant discrepancies between simulation and data in other distributions as a results of the rescaling.

\section{Predictions for the \protect\EIC}
\label{Sec:EIC}

\subsection{Diffractive DIS}
\label{Sec:EIC-DDIS}

For the analysis of diffractive DIS events we implemented a routine for \Rivet~\cite{Bierlich:2019rhm}, loosely based on the measurement in~\cite{H1:2015okx}, with the following phase space:
The photon virtuality was restricted to $4 \UGeV^2 < Q^2 < 110 \UGeV^2$ and we clustered jets in the lab frame using the $k_T$ algorithm with $R = 1$ within pseudorapditiy $\left| \eta \right| < 4$, demanding at least two jets with transverse energy $E_T$ of at least 5.5 and 4 GeV, respectively.
We assumed proton-tagging and reconstructed the $x_\Pom$ as
\begin{equation}
    x_\Pom = 1 - \frac{E_p^\prime}{E_p}
\end{equation}
which had to satisfy $x_\Pom < 0.1$, to allow for more phase space for jet production compensating for the lower beam energies.
The momentum transfer was restricted to $\left| t \right| < 0.6 \UGeV^2$ and we defined
\begin{equation}
    z_\Pom = \frac{Q^2 + M_{12}^2}{Q^2 + M_X^2} \ .
\end{equation}

\begin{figure}[h!]
    \centering
    \begin{tabular}{ccc}
        \includegraphics[width=.3\linewidth]{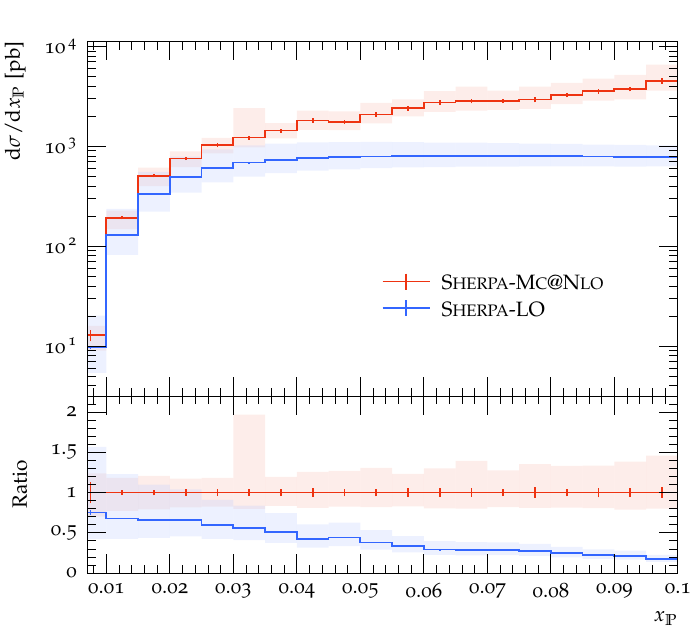} &
        \includegraphics[width=.3\linewidth]{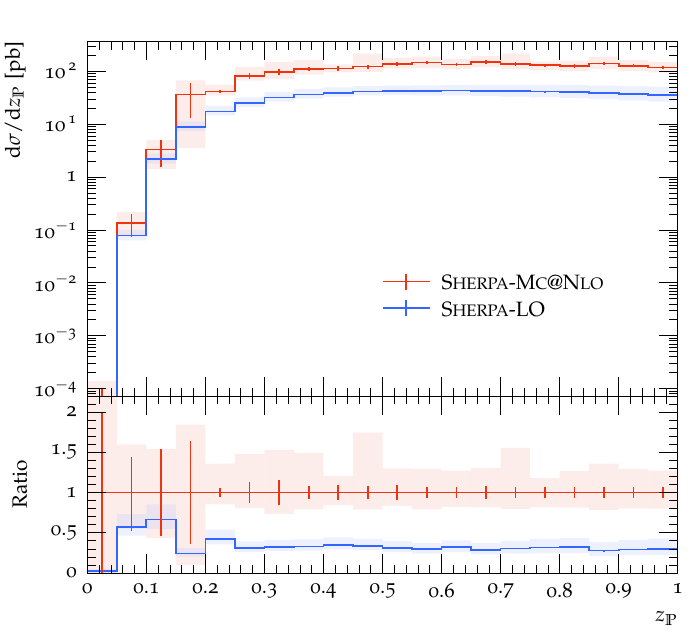} &
        \includegraphics[width=.3\linewidth]{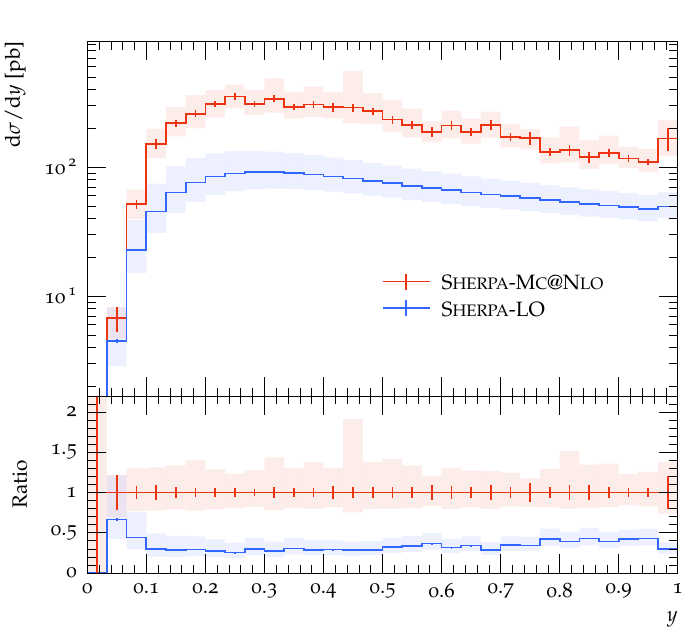} \\
        \includegraphics[width=.3\linewidth]{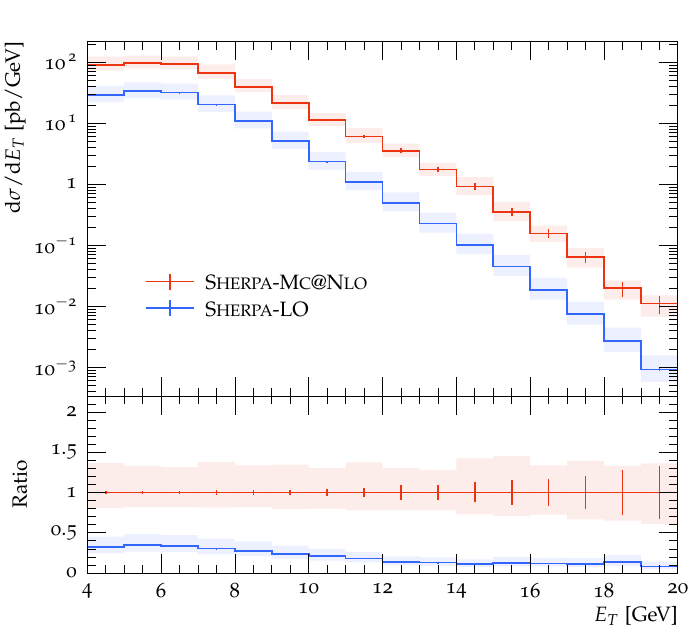} &
        \includegraphics[width=.3\linewidth]{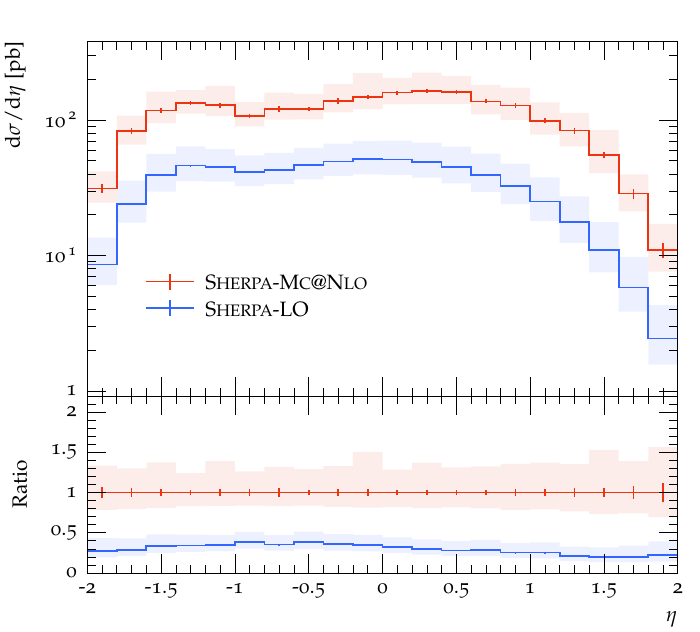} &
        \includegraphics[width=.3\linewidth]{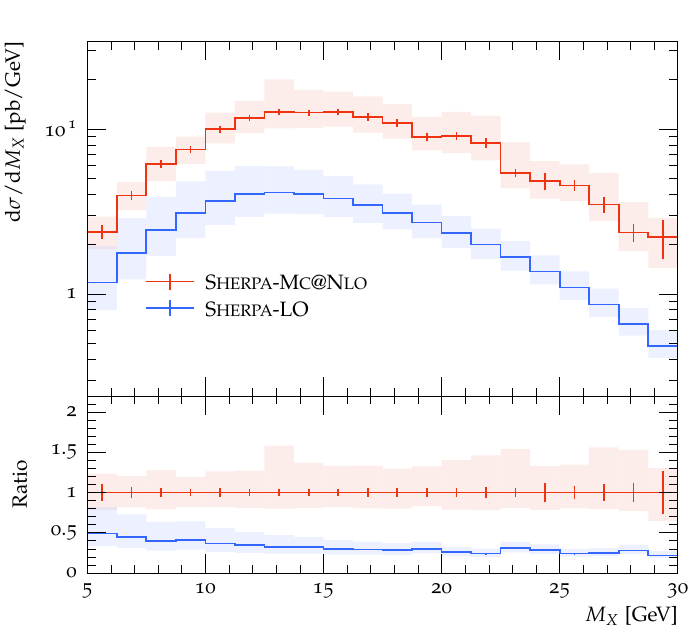} \\
        \includegraphics[width=.3\linewidth]{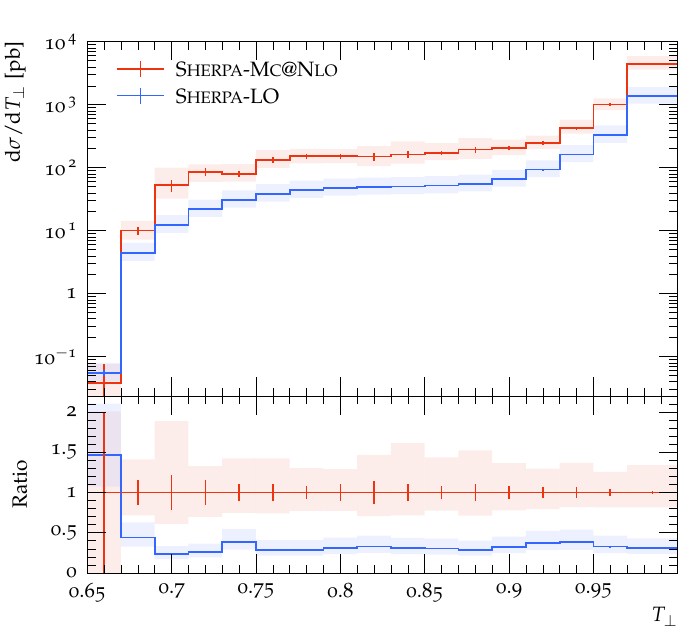} &
        \includegraphics[width=.3\linewidth]{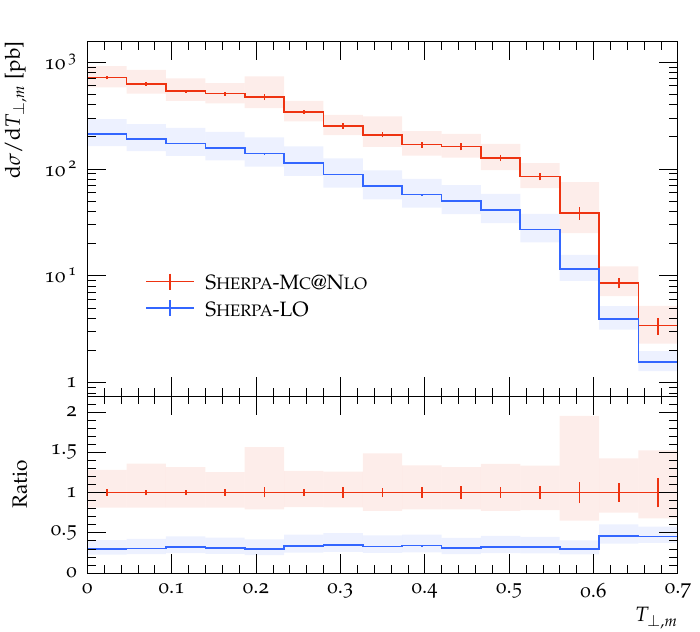} &
        \includegraphics[width=.3\linewidth]{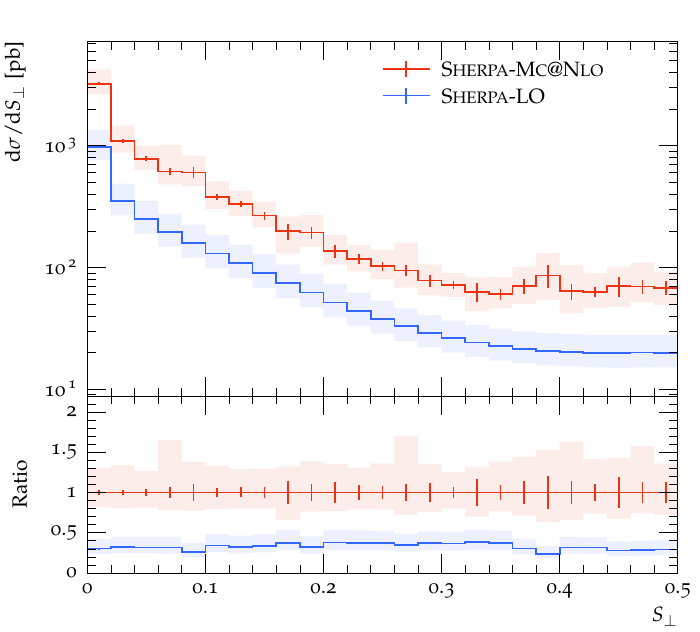}
    \end{tabular}
    \caption{Predictions at LO and \protect\MCatNLO accuracy of \protect\Sherpa for a range of observables relevant for diffractive DIS at the \protect\EIC, namely: the momentum ratios $x_\Pom$, $z_\Pom$ and inelasticity $y$ (upper row, left to right), leading jet transverse momentum $E_T^{\rm jet1}$, average jet rapidity $\langle\eta\rangle$ and diffractive mass $M_X$ (middle row, left to right), transverse thrust $T_\perp$, thrust minor $T_{\perp,m}$ and transverse sphericity $S_\perp$ (bottom row, left to right). }
    \label{fig:eic-ddis}
\end{figure}
In the upper two rows of Fig.~\ref{fig:eic-ddis} we exhibit the distributions in the momentum fractions $x_\Pom$ and $z_\Pom$ and the inelasticity $y$ and the distributions in the leading jet transverse energy $E_T^{\rm jet1}$, the average jet pseudorapidity $\langle\eta\rangle$ and the diffractive mass $M_X$.
We not, again, significant $K$-factors between LO and \MCatNLO predictions ranging up to values of about 5 in the forward regime, testaments to the lower energy scales of the processes we study here.
In addition to these observables we also display in the lower row of the same figure event shape distributions, in particular transverse thrust $T_\perp$, transverse thrust minor $T_{\perp m}$ and transverse sphericity $S_\perp$.
These observables are defined by
\begin{equation}
T_\perp = \mathrm{max}_{\vec n_{\mathrm{T}}} \frac{\sum_i \left| \vec p_{\mathrm{T},i} \cdot \vec n_{\mathrm{T}} \right|}{\sum_i \vec p_{\mathrm{T},i}}\;,\;\;\;
T_{\perp,m} = \mathrm{max}_{\vec n_{\mathrm{T}}} \frac{\sum_i \left| \vec p_{\mathrm{T},i} \times  \vec n_{\mathrm{T}} \right|}{\sum_i \vec p_{\mathrm{T},i}}\;,\;\;\;
S_\perp = \frac{2 \lambda_2}{\lambda_1 + \lambda_2}\;,
\end{equation}
and $n_{\mathrm{T}}$ is the transverse-thrust axis that maximises the $T_\perp$ and $\lambda_{1,2}$ are the eigenvalues of the transverse linearised sphericity tensor $\mathbf{S_{\alpha \beta}}$
\begin{equation}
    \mathbf{S_{\alpha \beta}} = \frac{1}{\sum_i \left| \vec p_{\mathrm{T},i} \right|} \sum_i \frac{1}{\left| \vec p_{\mathrm{T},i} \right|}
    \begin{pmatrix}
        p_{i,x}^2 & p_{i,x} p_{i,y} \\
        p_{i,y} p_{i,x} & p_{i,y}^2
    \end{pmatrix}.
\end{equation}

While the events are broadly dominated by dijet kinematics, the event shape distributions indicate a non-negligible contribution from three-jet events.
Clearly, the data taken at the \EIC will complement the \hera data in the high-$x$ region~\cite{Armesto:2021fws} in updated fits to the DPDFs.

\subsection{Diffractive photoproduction}
\label{Sec:EIC-DPHO}

\begin{figure}[h!]
    \centering
    \begin{tabular}{ccc}
        \includegraphics[width=.3\linewidth]{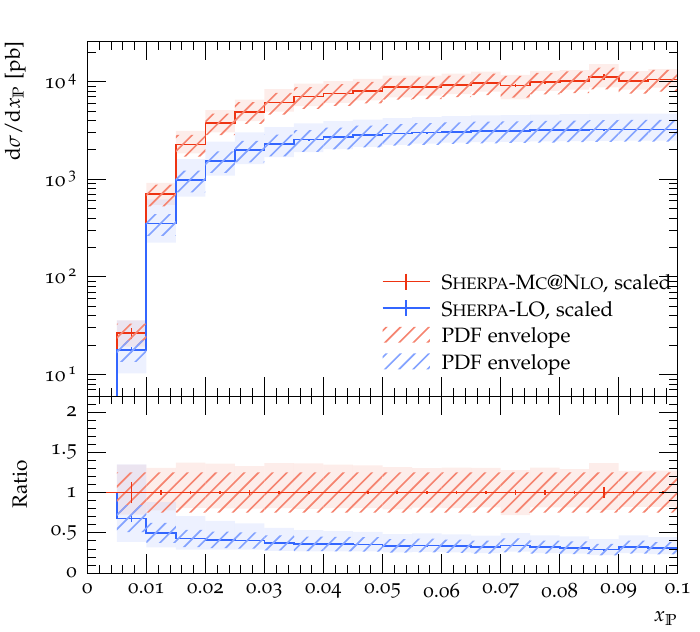} &
        \includegraphics[width=.3\linewidth]{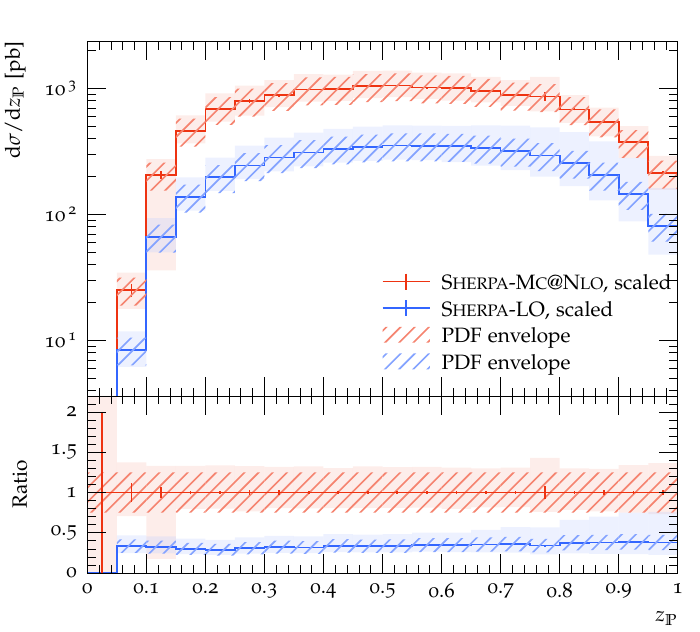} &
        \includegraphics[width=.3\linewidth]{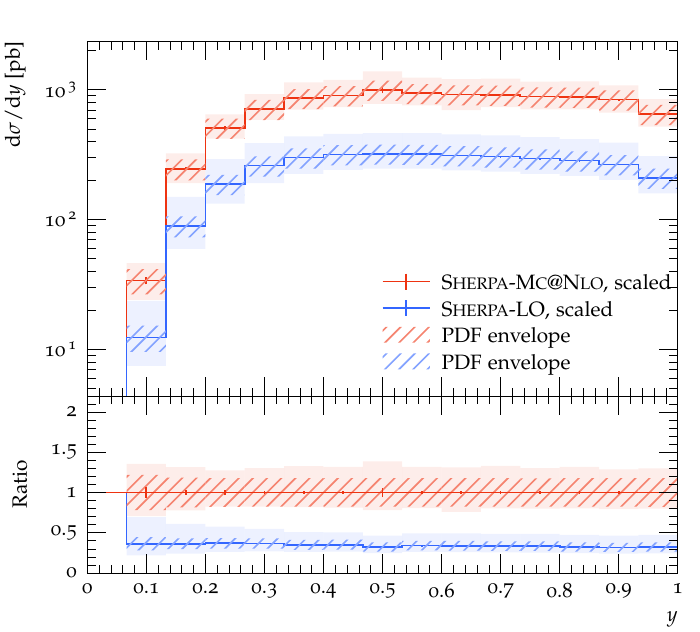} \\
        \includegraphics[width=.3\linewidth]{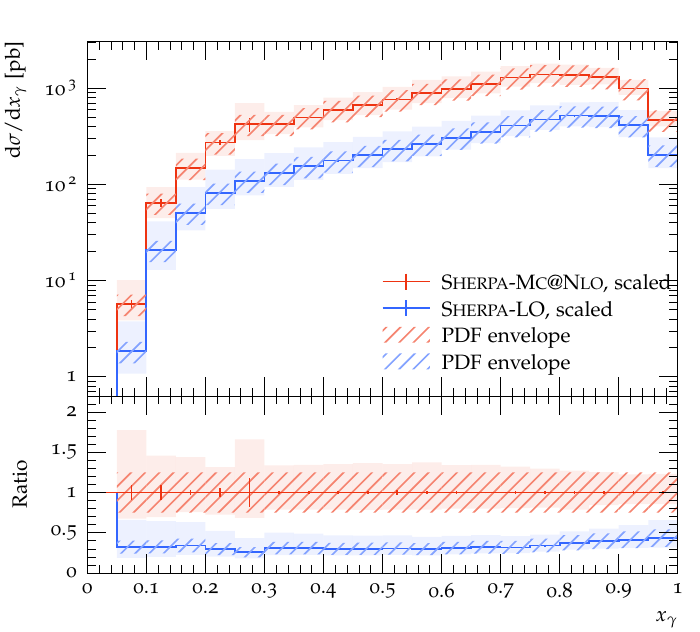} &
        \includegraphics[width=.3\linewidth]{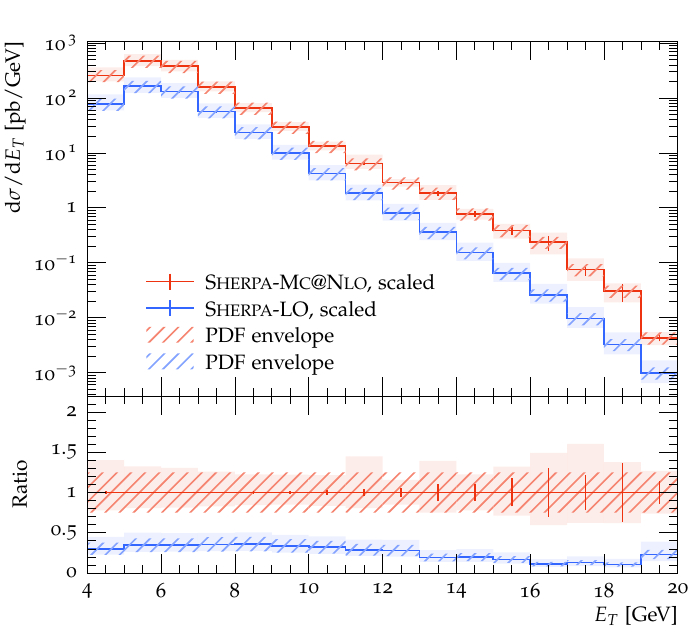} &
        \includegraphics[width=.3\linewidth]{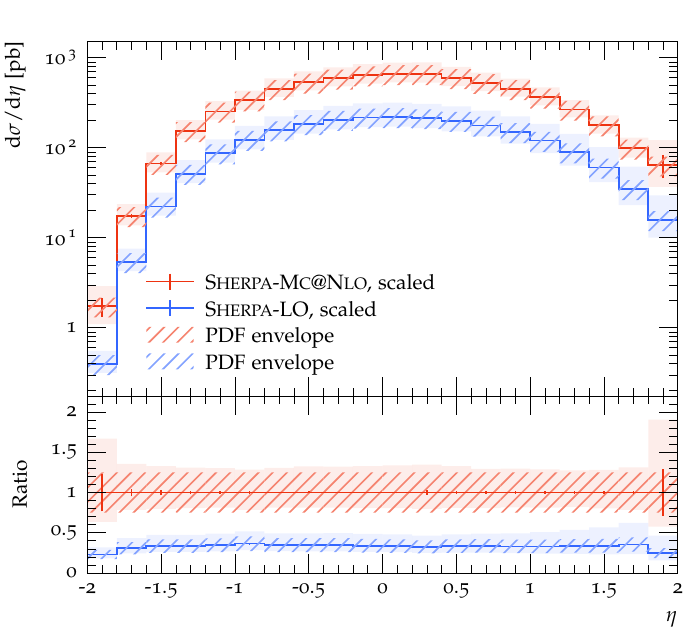} \\
        \includegraphics[width=.3\linewidth]{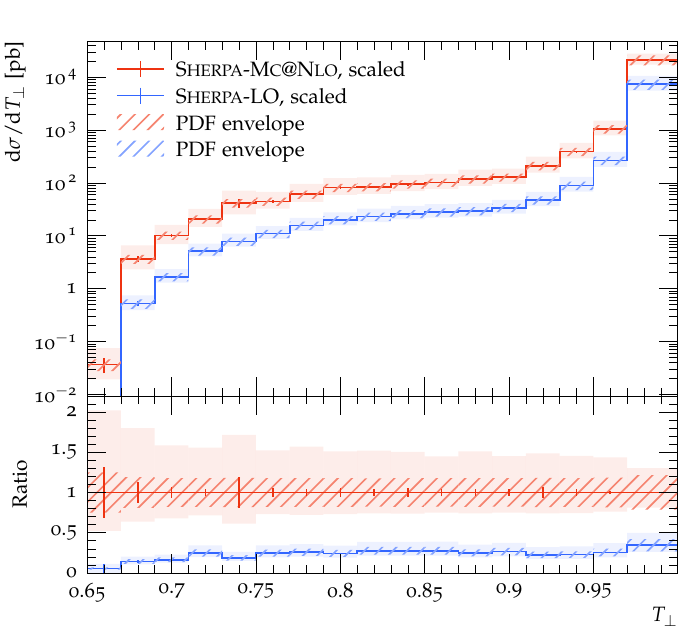} &
        \includegraphics[width=.3\linewidth]{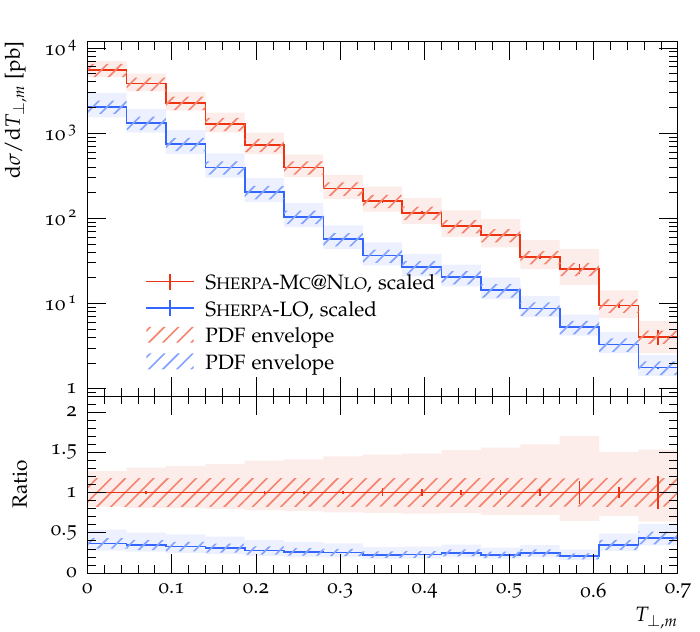} &
        \includegraphics[width=.3\linewidth]{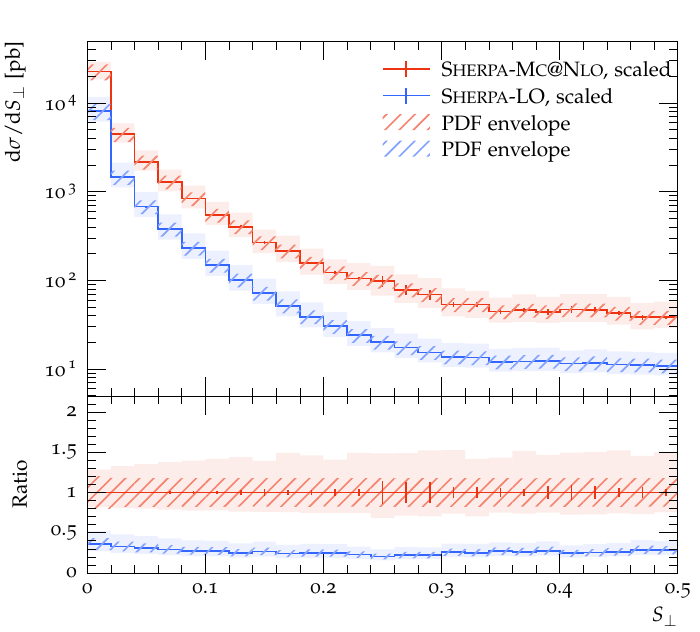}
    \end{tabular}
    \caption{Predictions at LO and \protect\MCatNLO accuracy of \protect\Sherpa of momentum ratio $x_\Pom$, $z_\Pom$ and $x_\gamma$ (upper row, left to right), of
    inelasticity $y$, leading jet transverse momentum $E_T^{\rm jet1}$ and average jet pseudorapidity $\langle\eta\rangle$ (middle row, left to right), and of event shapes transverse thrust $T_\perp$, transverse thrust minor $T_{\perp m}$ and transverse sphericity $S_\perp$ (bottom row, left to right) in diffractive photoproduction at the \protect\EIC. }
    \label{fig:eic-dpho}
\end{figure}

Measuring diffractive photoproduction at the \EIC will shed new light on factorisation breaking, and expand on some of the findings highlighted in Sec.~\ref{Sec:FactorisationBreaking}.
To obtain predictions for this process, we calculated an average over the suppression factors for direct and resolved photon processes from Tab.~\ref{tab:fitting-factors}: $R_\mathrm{res}^\mathrm{(EIC)} = 0.8 \pm 0.2$ and $R_\mathrm{dir}^\mathrm{(EIC)} = 0.4 \pm 0.1$ for the resolved and direct component, respectively.
Additionally to the scale uncertainty, we obtained obtained envelopes for the uncertainties from the suppression factors.
We used the same cuts and settings as for the predictions for diffractive DIS, see the previous subsection~\ref{Sec:EIC-DDIS}, restricting the virtuality to $Q^2 < 4 \UGeV^2$ and defined different momentum fractions as

\begin{equation}
    x_\Pom = 1 - \frac{E_p^\prime}{E_p}\;,\;\;\;
    z_\Pom = \frac{\sum_j E_T^{(j)} \mathrm{e}^{\eta^{(j)}}}{2 x_\Pom E_p}\;,\;\;\;
    y = 1 - \frac{E_e^\prime}{E_e}\;,\;\;\;
    x_\gamma = \frac{\sum_j E_T^{(j)} \mathrm{e}^{-\eta^{(j)}}}{2 y E_e}
\end{equation}
In Fig.~\ref{fig:eic-dpho}, we display the momentum fractions, inelasticity and jet observables as well as event shapes in diffractive photoproduction.
Uncertainties due to the fitting to factorisation breaking and due to scale choices are of comparable size, and the $K$ factors between the \MCatNLO and LO accuracy again reach values of about 5.
The event shapes exhibit an anticipated effect, namely that the limited phase space leads to lower multiplicities and an even stronger dominance of dijet events compared to the DDIS events in the previous chapter, manifest in the sharper peak at low values of $1-T_\perp$, $T_{\perp m}$, and $S_\perp$.

\clearpage
\section{Summary}\label{Sec:summary}

Diffraction played an important role at \hera{}, making up 10\% of the total cross section. 
We simulated (hard) diffractive jet production at Next-to-Leading-Order in \Sherpa for electron-proton collision in both the DIS and photoproduction regimes, matching our calculation to the parton shower.
This results in the first fully differential hadron-level calculation of hard diffraction at \MCatNLO accuracy and provides an important accuracy standard for future \EIC predictions.

Validating our simulation against data from the \hone and \zeus collaborations we see excellent agreement for Diffractive DIS setups. 
In the Diffractive Photoproduction regime, we observe significant discrepancies compared to the data, confirming the findings in previous Fixed Order calculations. 
We review and discuss different ans\"atze to explain the differences in view of our hadron-level simulations and conclude that none of the proposed solutions suffices to conclusively clarify the mechanism of factorisation breaking in this regime.
We argue that the factorisation breaking happens in fact in {\em both} the direct and resolved component in diffractive photoproduction. 
A coherent mechanism would need to take into account the suppression of the real correction to the "direct" component at NLO. 
We quantify the suppression in this component by fitting the two components to the data.

Lastly, we presented predictions for Diffractive DIS and Diffractive Photoproduction at the \EIC, where for the latter we estimated the suppression due to factorisation breaking by means of the fits to the \hone and \zeus data.
Data taking at the \EIC will provide more insights into the exact mechanism of factorisation breaking and a thorough comparison to theory predictions will determine the exact nature of the corresponding mechanism. 
With this understanding it should be possible to apply this to hadron colliders, hence allowing to study this phenomenon on the basis of the vast data taken at the \lhc{}.

\section*{Acknowledgements}
We are indebted to our colleagues in the \Sherpa collaboration, for numerous discussions and technical support.
F.K.\ gratefully acknowledges funding as Royal Society Wolfson Research fellow.
F.K.\ and P.M.\ are supported by STFC under grant agreement ST/P006744/1.

\bibliographystyle{unsrt}
\bibliography{refs}
\end{document}